\documentclass[aps,xpre,longbibliography,showpacs,amsmath,amssymb]{revtex4-1}
\usepackage{amsmath,amssymb,latexsym,epsfig,graphics,epsf,bm}
\usepackage{hyperref}
\usepackage{graphics}
\usepackage{gensymb}
\usepackage{float}
\usepackage{footnote}
\usepackage{tabularx,booktabs}
\usepackage{subfigure}
\usepackage[toc,page]{appendix}
\usepackage[dvipsnames]{xcolor}

\newcommand{\be}{\begin{equation}}
\newcommand{\ee}{\end{equation}}
\newcommand{\fig}[1]{Fig.~\ref{#1}}
\newcommand{\Fig}[1]{Figure~\ref{#1}}
\newcommand{\eq}[1]{Eq.~(\ref{#1})}

\newcommand{\Sex}{{S}_{\rm ex}}
\newcommand{\tSex}{{\tilde S}_{\rm ex}}

\newcommand{\bF}{\bold F}
\newcommand{\tbF}{\tilde{\bold F}}
\newcommand{\bR}{\bold R}
\newcommand{\bRet}{{\bf R}_{\rm 1}}
\newcommand{\bRto}{{\bf R}_{\rm 2}}
\newcommand{\bRa}{{\bf R}_{\rm a}}
\newcommand{\bRb}{{\bf R}_{\rm b}}
\newcommand{\br}{\bold r}
\newcommand{\tbR}{{\tilde{\bold R}}}
\renewcommand{\tt}{{\tilde t}}

\begin{document}
	
	\title{Scaling properties of liquid dynamics predicted from a single configuration: Small rigid molecules}
	\date{\today}
	\author{Zahraa Sheydaafar}
	\author{Jeppe C. Dyre}
	\author{Thomas B. Schr{\o}der}\email{tbs@ruc.dk}
	\affiliation{Glass and Time, IMFUFA, Department of Science and Environment, Roskilde University, P.O. Box 260, DK-4000 Roskilde, Denmark}
	
\begin{abstract}
Isomorphs are curves in the thermodynamic phase diagram along which structure and dynamics are invariant to a good approximation. There are two main ways to trace out isomorphs, the configurational-adiabat method and the direct-isomorph-check method. Recently a new method based on the scaling properties of forces was introduced and shown to work very well for atomic systems [T. B. Schr{\o}der, \textit{Phys. Rev. Lett.} \textbf{2022}, \textit{129}, 245501]. A unique feature of this method is that it only requires a single equilibrium configuration for tracing out an isomorph. We here test generalizations of this method to molecular systems and compare to simulations of three simple molecular models: the asymmetric dumbbell model of two Lennard-Jones spheres, the symmetric inverse-power-law dumbbell model, and the Lewis-Wahnstr\"{o}m o-terphenyl model. We introduce and test two force-based and one torque-based methods, all of which require just a single configuration for tracing out an isomorph. Overall, the method based on requiring invariant center-of-mass reduced forces works best. 
\end{abstract}

\maketitle

\section{Introduction}
Isomorphs are curves of invariant structure and dynamics in the thermodynamic phase diagram \cite{IV,dyr18}. Such curves exist in systems that have strong correlations between the constant-volume canonical-ensemble equilibrium fluctuations of virial and potential energy \cite{IV,dyr14}, a characteristic property of so-called R-simple (``strongly correlating'') systems \cite{ing12b,bai13,fle14,Pra14}. The Pearson correlation coefficient $R$ between the equilibrium fluctuations of virial $W$ and potential energy $U$ is given by (where sharp brackets denote $NVT$ canonical averages and $\Delta$ denotes the deviation from the equilibrium mean value, e.g., $\Delta U \equiv U - \langle U \rangle$):

\be\label{R}
R = \dfrac{\langle \Delta W \Delta U \rangle}{\sqrt{\langle (\Delta W)^2 \rangle \langle (\Delta U)^2 \rangle}}\,.
\ee
For an inverse-power-law (IPL) system with pair potential proportional to $r^{-n}$ ($r$ is the pair distance) the correlation is perfect ($R = 1$) because $W = (n/3) U$ for all configurations. Slightly less than perfect correlation still leads to almost invariant structure and dynamics along the so-called isomorphs (see below), and the class of R-simple liquids is usually delimited by the pragmatic criterion $R>0.9$.  

Invariance of structure and dynamics along isomorphs occurs when \textit{reduced} units are applied that, notably, depend on the thermodynamic state point in question. In this unit system, the length unit $l_0$ is defined from the particle-number density $\rho\equiv N/V$ where $N$ is the particle number and $V$ the system volume, the temperature $T$ defines the energy unit $e_0$, and the density and thermal velocity define the time unit $t_0$. Specifically, if $m$ is the average particle mass, reduced units are defined by \cite{IV,I,ing12b} by

\be\label{units}
l_0=\rho^{-1/3}\,\,,\,e_0=k_BT\,\,,\,t_0=\rho^{-1/3}\sqrt{m/k_BT}\,.
\ee 
Isomorph theory has been applied successfully to different classes of systems \cite{bel21}, including simple atomic systems in both the liquid and solid phases \cite{I,II,sch09,pal14,EXPII,hey15b,cos16,mau18}, molecular systems \cite{ing12b,kop20}, and the 10-bead Lennard-Jones chain \cite{vel14}. Isomorph-theory predictions have also been verified in experiments on glass-forming van der Waals molecular liquids \cite{xia15,han18}.

\begin{figure}[htbp!]
	\centering
	\includegraphics[width=8cm]{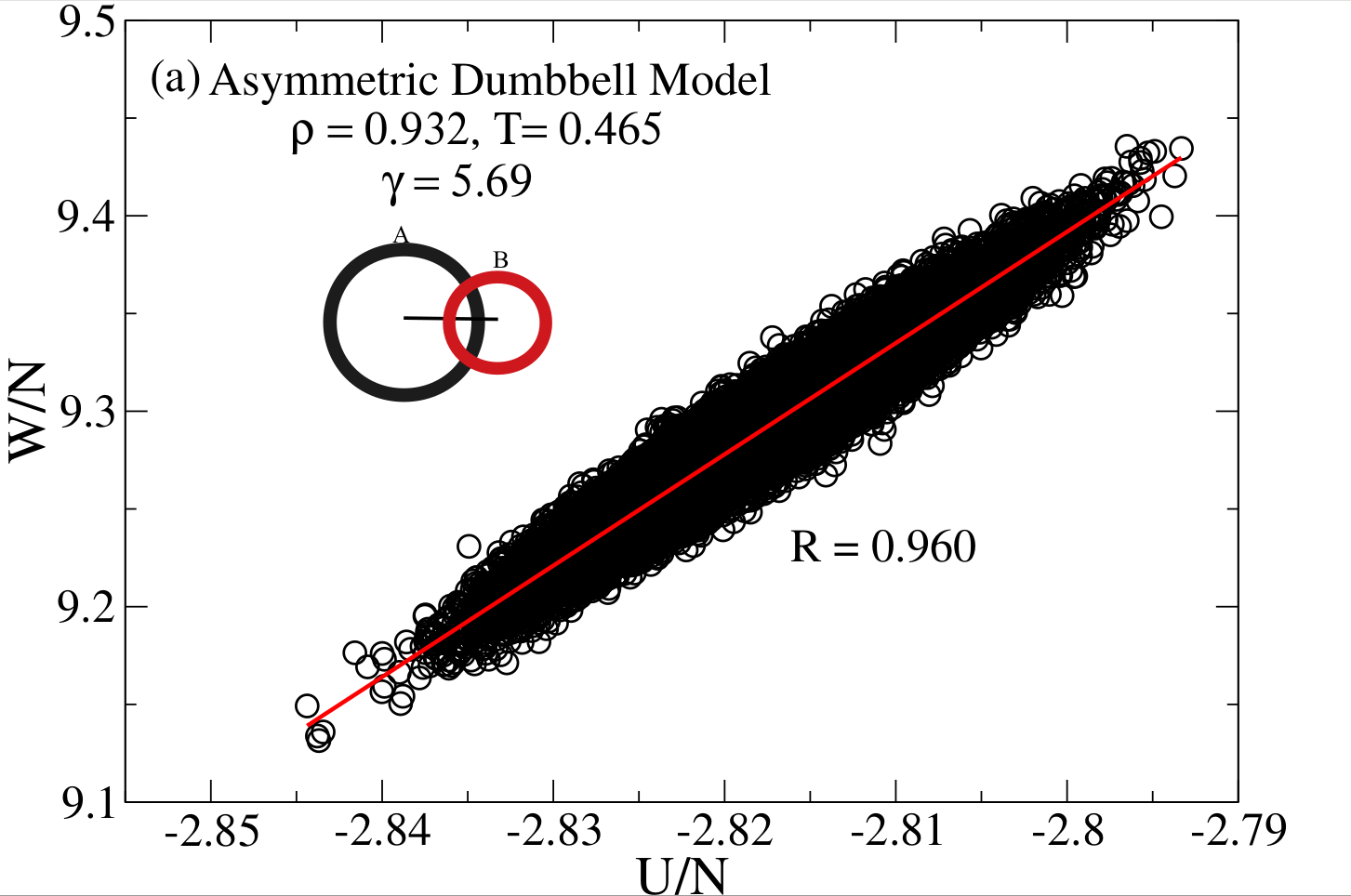}
	\includegraphics[width=8cm]{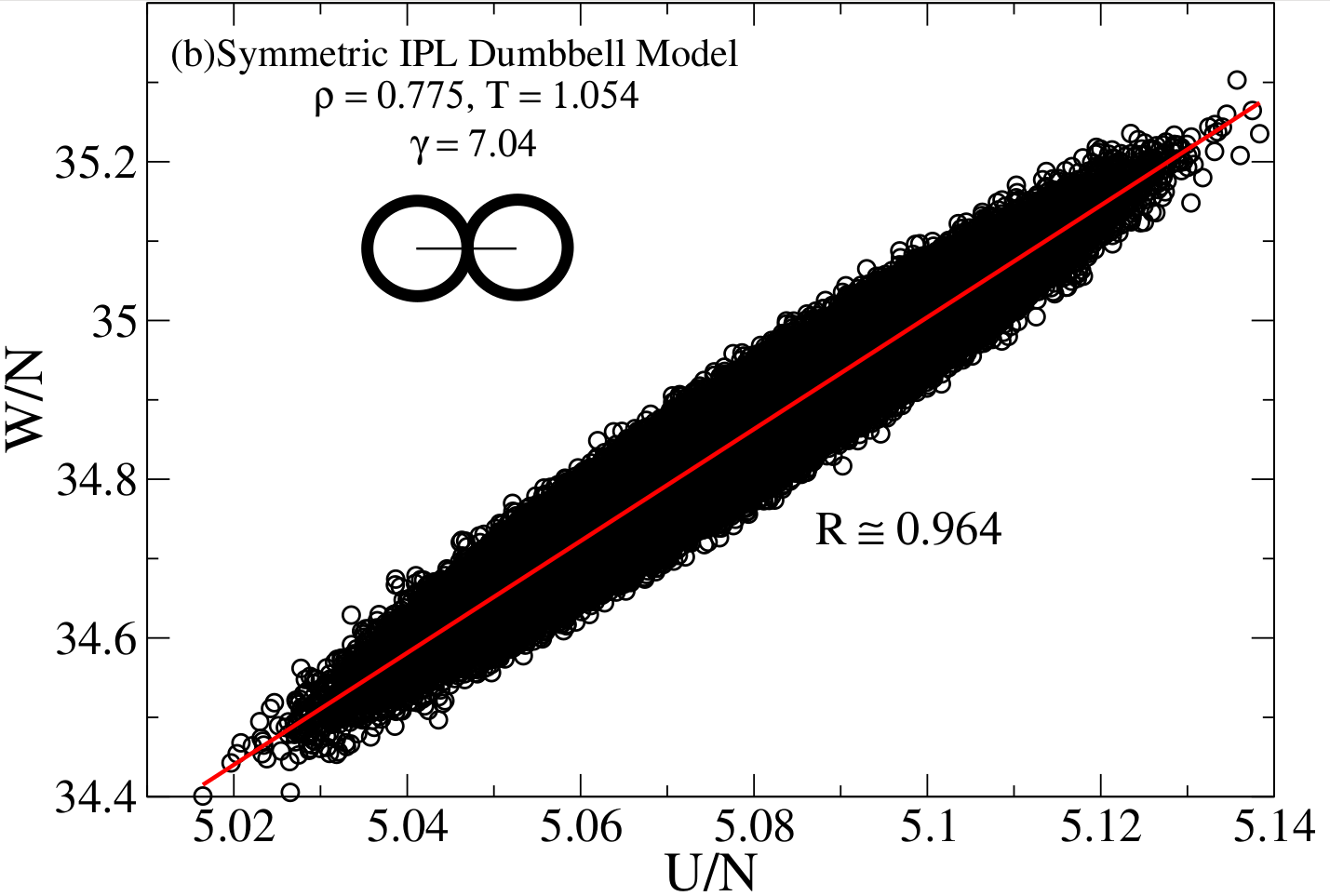}
	\includegraphics[width=8cm]{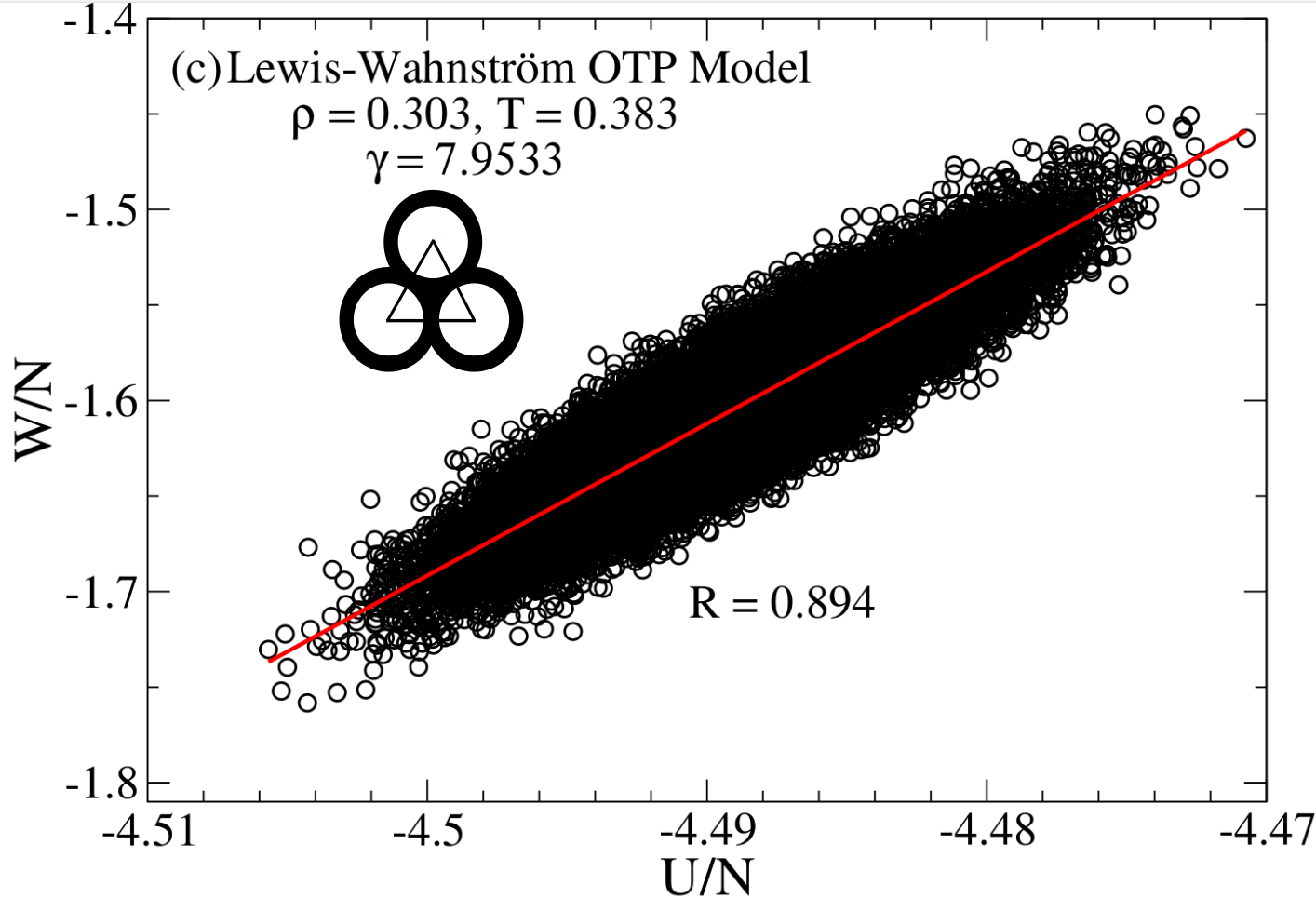}
	\caption{\label{FIG1} Scatter plots of equilibrium fluctuations of virial $W$ and potential energy $U$ for 
        (a) the asymmetric dumbbell model, 
        (b) the symmetric IPL dumbbell model, and 
        (c) the Lewis-Wahnstr\"{o}m OTP model. 
        All three models have bonds defined by rigid constraints; model details are given in Sec. II. The state points simulated here are used in the following as reference state points for tracing out isomorphs of the three models. 
 }   
\end{figure}

In 2012 Ingebrigtsen \textit{et al.} studied numerically isomorphs for liquid molecular systems composed of small rigid molecules \cite{ing12b}. This publication found isomorphs for the asymmetric dumbbell model (\fig{FIG1}(a)), the symmetric inverse-power-law (IPL) dumbbell model (\fig{FIG1}(b)), and the Lewis-Wahnstr\"{o}m o-terphenyl (OTP) model (\fig{FIG1}(c)). Reference \onlinecite{ing12b} used the so-called configurational-adiabat method to trace out isomorphs. To explain this method, consider a scatter plot of virial versus potential energy of configurations taken from an equilibrium simulation (\fig{FIG1}). The linear-regression slope of such a plot, which is termed the density-scaling exponent $\gamma$, is given \cite{I,fra11a,Kop19,fra19} by the following general statistical-mechanics identity: 

\begin{equation} \label{gamma}
\gamma 
= \left( \frac{\partial \ln T}{\partial \ln \rho}\right)_{\mathrm{S_{ex}}}
\equiv \frac{\langle \Delta U \Delta W \rangle}{\langle (\Delta U)^2 \rangle} \,. 
\end{equation}
We remind the reader that the excess entropy $\Sex$ is the entropy minus the ideal-gas entropy at same density and temperature. For all systems, whether they are R-simple or not, \eq{gamma} allows one to trace out configurational adiabats, i.e., curves in the phase diagram of constant $\Sex$. This is done by calculating the two canonical averages in \eq{gamma} at an initial ``reference'' state point, changing density slightly, e.g., by 1\%, and from \eq{gamma} calculating the change in temperature that will keep $\Sex$ constant. At the new state point the canonical averages are recalculated, and so on. For R-simple liquids, isomorph invariance of $\Sex$ follows from the isomorph invariance of structure \cite{ing12b,sch09,dyr16,dyr18}; thus for an R-simple system the configurational-adiabat method in principle traces out isomorphs correctly. 

A much faster method for generating isomorphs termed the ``direct isomorph check'' and works as follows \cite{IV}. It can be shown that two configurations of an R-simple system, which can be scaled uniformly into one another, belong to state points on the same isomorph and have proportional Boltzmann factors, i.e., \cite{IV}

\be\label{Boltz}
e^{-U(\bRet)/k_{\mathrm{B}}T_1} = C_{12} e^{-U(\bRto)/k_{\mathrm{B}}T_2}.
\ee
Here, if $\bR\equiv (\br_1,...,\br_N)$ is the configuration of all $N$ particles, $\bRet$ and $\bRto$ are two configurations of (particle) density $\rho_1$ and $\rho_2$, respectively, obeying $\bRto=\left(\rho_1/\rho_2 \right)^{1/3}\bRet$, and $C_{12}$ is a constant that depends only on the two state points in question, not on the configurations. By taking the logarithm of \eq{Boltz} we get

\be\label{eq:DIC}
U(\bRto) = \dfrac{T_2}{T_1} U(\bRet) + k_{\mathrm{B}}T_2 \ln C_{12}.
\ee
Thus selecting configurations $\bRet$ from an equilibrium NVT simulations at the state point $(\rho_1,T_1)$ and plotting $U(\bRto)$ versus $U(\bRet)$ in a scatter plot results in a strong correlation if the system is R-simple, and $T_2$ can be calculated from the slope of this scatter plot.

Recently a new method for tracing out isomorphs was introduced and demonstrated to work very well for two atomic systems, the viscous Kob-Andersen binary Lennard-Jones liquid and the single-component Lennard-Jones liquid \cite{sch22}. As detailed below, the new method is based on the scaling of forces upon uniform scaling of configurations. A unique feature is that this method can be applied by considering just one configuration. In the present paper we discuss three generalizations of this method to molecular systems and test them on the molecular systems introduced above.

\section{Simulation details}\label{appendix}
	
We studied three molecular systems with rigid bonds, the asymmetric dumbbell (ASD) model \cite{vra01,gal07,cho10a,cho10b,ing12b,fra17,san18,dom20} (for N = 5000 atoms), the symmetric $r^{-18}$ IPL dumbbell model (N = 5000), and the Lewis-Wahnstr\"{o}m OTP model \cite{otp2} (N = 3000). All three models are known to have good isomorphs \cite{ing12b}.
	
An ASD molecule consists of a rigidly bonded large (A) and small (B) Lennard-Jones (LJ) particle \cite{vra01,sch09}. The length of the bond is $0.584$ in the LJ units defined by the large particle ($\sigma_{AA} = 1$, $\epsilon_{AA} = 1$, and $m_A = 1$). The parameters of the ASD model ($\sigma_{AB} = 0.894$, $\sigma_{BB} = 0.788$, $\epsilon_{AB} = 0.342$, $\epsilon_{BB} = 0.117$, $m_B = 0.195$) were chosen to mimic toluene \cite{sch09}. The intermolecular interactions are given by the LJ pair potential,
\be
v_{ij}(r_{ij}) = 4 \epsilon_{\alpha\beta} \left[\left(\dfrac{\sigma_{\alpha\beta}}{r_{ij}}\right)^{12} - \left(\dfrac{\sigma_{\alpha\beta}}{r_{ij}}\right)^{6}\right]. 
\ee
Here $r_{ij}$ is the distance between particles $i$ and $j$, and $\alpha,\beta\in \{A,B\}$ are the types of particles $i$ and $j$ respectively.   

The symmetric IPL dumbbell model consists of two identical particles connected by a rigid bond, also of length $0.584$ in LJ units. The intermolecular interactions are here given by the inverse power-law (IPL) pair potential,
\be
v(r) = \epsilon \left(\dfrac{\sigma}{r}\right)^{18}\,.
\ee
The two parameters and particle masses were set to unity. 

Finally, the Lewis-Wahnstr\"{o}m OTP model consists of three identical LJ particles, which are connected by rigid bonds in an isosceles triangle with unit side length and top angle $75\degree$. Again, the LJ parameters were set to unity. 

The Molecular Dynamics simulations were performed in the $NVT$ ensemble with a Nose-Hoover thermostat using RUMD (\url{http://rumd.org}), which is an open-source package optimized for GPU computing \cite{RUMD}. Details of how to calculate the virial of a constrained system \cite{edb86} are given in Appendix A of Ref. \onlinecite{ing12b}.

\section{Three single-configuration methods for tracing out isomorphs}

While straightforward, generating isomorphs by means of \eq{gamma} requires many equilibrium configurations. As shown recently for atomic systems, good statistics can be obtained, however, from the scaling properties of the forces of a single configuration \cite{sch22}. The idea is to make use of the fact that forces in reduced units are isomorph invariant. To show this we refer to the basic equation of isomorph theory \cite{sch14},

\be\label{Urel}
U(\bR)
\,=\,U(\rho,\Sex(\tbR))\,.
\ee
Here $U(\rho,\Sex)$ is the thermodynamic average potential energy at the state point with density $\rho$ and excess entropy $\Sex$, $\tbR\equiv\bR/l_0=\rho^{1/3}\bR$ is the reduced configuration vector, and $\Sex(\tbR)$ is the microscopic excess-entropy function defined in Ref. \onlinecite{sch14}. The fact that this function only depends on a configuration's \textit{reduced} coordinates is a consequence of the hidden scale invariance condition
, $U(\bRa)<U(\bRb) \Rightarrow U(\lambda\bRa)<U(\lambda\bRb)$, in which $\lambda$ is a uniform scaling parameter \cite{sch14}; this condition is equivalent to the system in question having strong virial potential-energy correlations, i.e., being R-simple \cite{I,sch14}.

It follows from \eq{Urel} that the $3N$-dimensional vector of forces on the particles, $\bF\equiv (\bF_1,...,\bF_N)$, is given by

\be\label{Frel}
\bF(\bR)
=-\nabla U(\bR)
= -\left(\frac{\partial U}{\partial\Sex}\right)_\rho \rho^{1/3}\tilde{\nabla}\,\,\Sex(\tbR)\,.
\ee
Since $(\partial U/\partial\Sex)_\rho=T$ this implies that the reduced force vector $\tbF \equiv l_0\bF/e_0={\rho^{-1/3}\bF}/{k_BT}$ is given by (in which $\tSex\equiv \Sex/k_B$)

\be\label{N2_red}
\tbF
=-\tilde{\nabla}\tSex(\tbR)\,.
\ee
The fact that $\tbF$ depends only on the \textit{reduced} coordinates means that the reduced-unit version of Newton's second law, $d^2\tbR/d\tt^2=\tilde{m}\,\tbF(\tbR)$ \cite{IV}, has no reference to the state-point density. This implies invariant reduced-unit dynamics along isomorphs defined as configurational adiabats, i.e., by $\Sex=$ Const. \cite{sch14}.

Given a reference state point $(\rho_1, T_1)$ and a different density, $\rho_2$, we now derive the equation for calculating the temperature $T_2$ with the property that the state point $(\rho_2, T_2)$ is on the same isomorph as $(\rho_1, T_1)$. If $\bR_1$ is a configuration taken from an equilibrium simulation of the reference state point and $\bR_2$ is the same configuration scaled uniformly to density $\rho_2$, i.e., $\bRto = \left(\rho_1/\rho_2 \right)^{1/3} \bRet$, it follows from \eq{N2_red} that because $\tbR_1=\tbR_2$ the reduced forces of the two configurations are identical,

\be\label{eq:F_inva}
\tbF(\bRet) 
= \tbF(\bRto)\,.
\ee
In particular, this implies that the lengths of the $3N$-dimensional reduced force vectors are identical, 
$|\tbF(\bRet)| = |\tbF(\bRto)|$. Since $\tbF \equiv \rho^{-1/3}\bF/{k_BT}$ this means that $T_2$ can be determined from

\be\label{eq:pre_temp}
T_2 = \frac{|\bF(\bRto)|}{|\bF(\bRet)|} \left(\frac{\rho_1}{\rho_2}\right)
^{1/3} T_1.
\ee

In the case of perfect scaling, i.e., if \eq{Urel} is fulfilled exactly, \eq{eq:pre_temp} ensures that \eq{eq:F_inva} is obeyed and thus that structure and dynamics are invariant in reduced units \cite{sch22}. However, the scaling is rarely perfect; in particular it is never perfect for systems with both repulsive and attractive forces. The choice made in Ref. \onlinecite{sch22} was to use \eq{eq:pre_temp} to get a value for $T_2$ and then use \eq{eq:F_inva} to estimate whether the scaling works well. When \eq{eq:F_inva} is not fulfilled exactly, the prediction of invariant reduced structure and dynamics becomes an approximation, and simulations must be performed to test its validity. This was done in Ref. \onlinecite{sch22} for two atomic systems where the method was demonstrated to work very well. In particular for the viscous Kob-Andersen binary Lennard-Jones mixture, the new force-based method outperforms the direct-isomorph-check method despite being based on information from just a single configuration.   

\begin{figure}[htbp!]
	\centering
	\includegraphics[width=7cm]{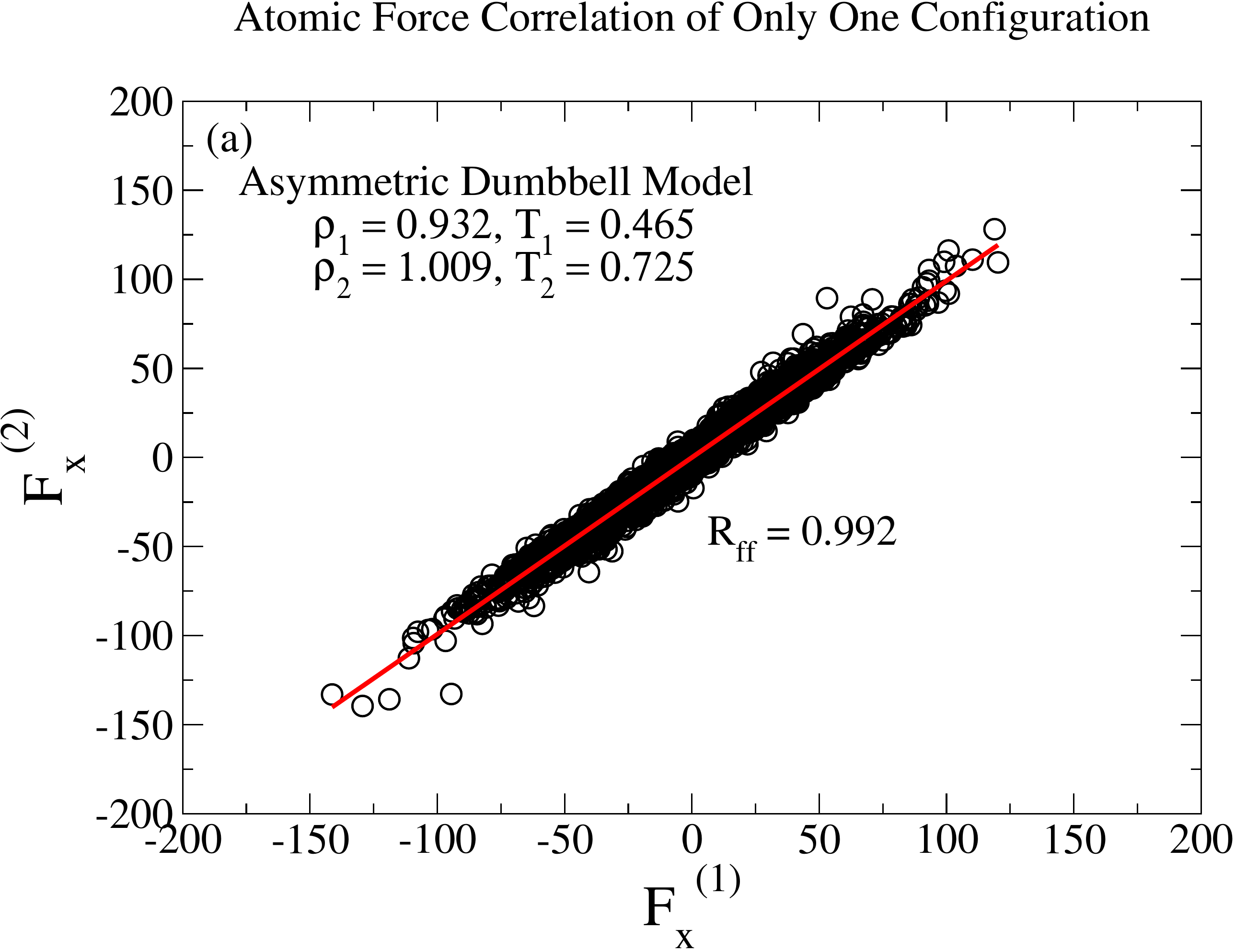}
	\includegraphics[width=7cm]{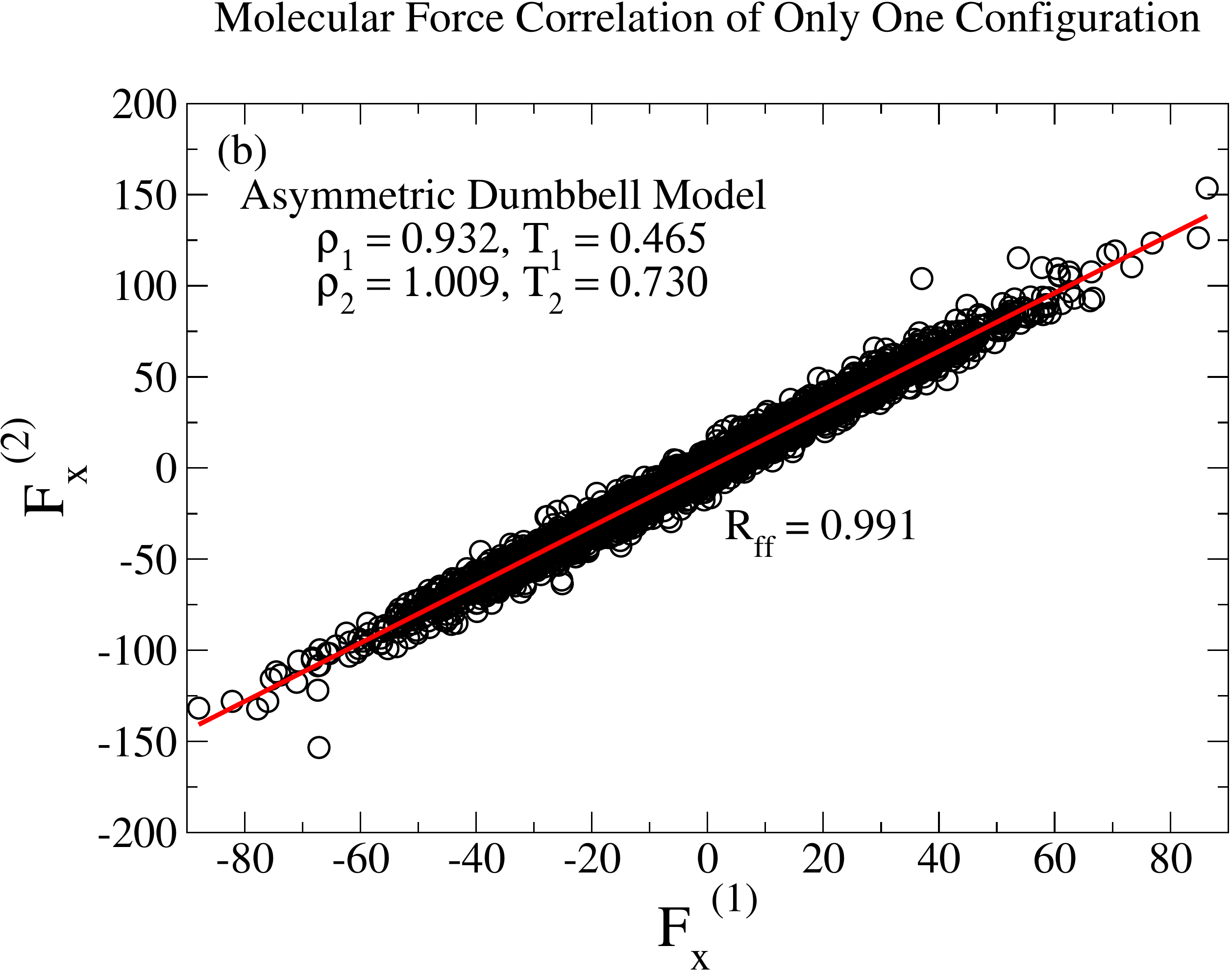}
	\includegraphics[width=7cm]{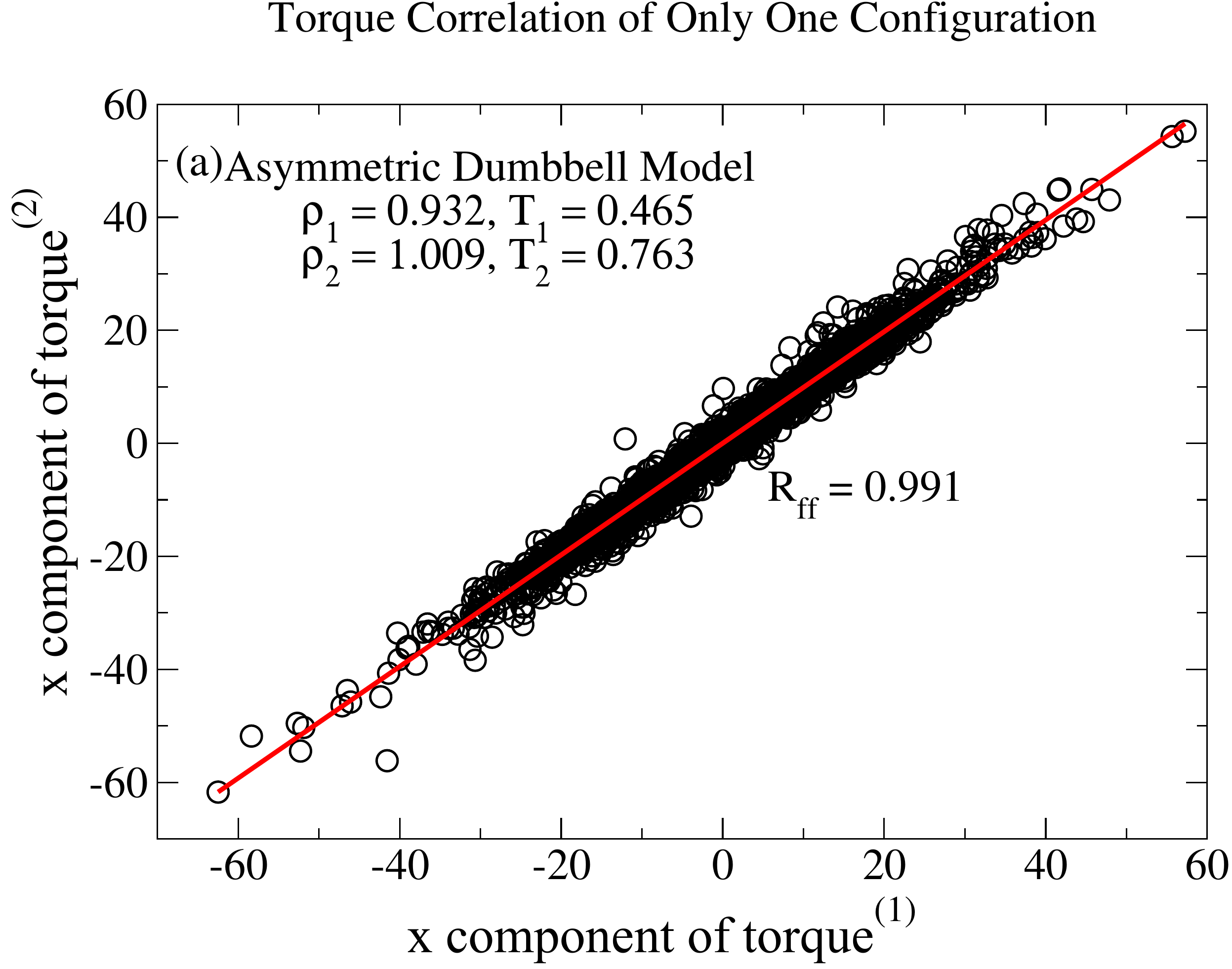}	
	\caption{\label{FIG4} Scaling plots of the three single-configuration methods for tracing out isomorphs illustrated by data for the      ASD model.
	(a) [``Atomic-force method''] shows a plot of all particle forces in one axis direction for a single configuration $\bR_1$ of the      reference state point $(\rho_1, T_1) = (0.932, 0.465)$ versus those of its uniformly scaled version to density $\rho_2=1.009$, i.e., $\bR_2=(\rho_1/\rho_2)^{1/3}\bR_1$. From  \eq{eq:pre_temp} one finds $T_2=0.725$.
	(b) [``Molecular-force method''] shows a similar plot based on the center-of-mass forces between the molecules; the slightly different $T_2=0.730$ is arrived at using this method. 
    (c) [``Torque Method''] shows data for for the torque of the molecules. Despite the strong correlation, the temperature $T_2$ identified by this method via \eq{eq:pre_tor} is quite different from those of the two force-based methods ($T_2 = 0.763)$.
		}
\end{figure}

How should one adopt the force method, which was devised for systems of point particles, to molecules with rigid bonds? One option is to do as described above, interpreting $\mathbf R$ as the vector of the positions of all atoms with $\bF$ as the vector of all corresponding forces. Performing the uniform scaling $\bRto = \left(\rho_1/\rho_2 \right)^{1/3} \bRet$ on the atomic level would violate the molecular bond-length constraint, however. Moreover, this approach would take into account intramolecular forces, the role of which are merely to ensure the rigid-bond constraint. In view of these issues, it makes more sense to focus on the forces on the molecules' center-of-masses. In that case we let $\mathbf R$ be the vector of all center-of-mass positions; the corresponding $\bF$ vector is then the vector of the resulting forces on each molecule from the surrounding molecules, and the scaling is applied to the center-of-masses keeping the bond lengths and orientations fixed. We refer to the center-of-mass forces as ``molecular forces'' and to this version of the force method as the ``molecular-force method''. In the tests presented below we include for comparison, however, the somewhat unphysical ``atomic-force method'' (using also center-of-mass scaling). 

We consider also a third method based on torques. Assuming again perfect scaling, the reduced-unit torque on each molecule is the same along an isomorph before and after uniform scaling, i.e., $\boldsymbol{\tilde{\tau}_1} = \boldsymbol{\tilde{\tau}_2}$ where $\boldsymbol{\tau}$ is the vector of all torques. Since the reduced-unit torque is defined by $\boldsymbol{\tilde{\tau}} \equiv {\boldsymbol{\tau}}/{e_0} = {\boldsymbol{\tau}}/{k_B T}$, this invariance implies that $T_2$ is given by

\begin{equation}\label{eq:pre_tor}
T_2 = \frac{\boldsymbol{|\tau_2|}}{\boldsymbol{|\tau_1|}} T_1\,.
\end{equation}

The three methods are illustrated in \fig{FIG4} for a 7\% density change applied to the ADP system, where (a) shows the x-coordinates of the forces on all particles plotted against the same quantities of the center-of-mass uniformly scaled configurations. \Fig{FIG4}(b) shows the same for the center-of-mass molecular forces that have no intramolecular contributions. We find a strong correlation between scaled and unscaled forces in both cases, but a slightly different prediction for $T_2$, which is $0.725$ using the atomic-force method and $0.730$ using the molecular-force method. \Fig{FIG4}(c) shows the corresponding correlation of torque components before and after scaling, showing again a high correlation but leading in this case to the significantly higher temperature $T_2 = 0.763$.

\begin{figure}[htbp!]
	\centering
	\includegraphics[width=5.4cm]{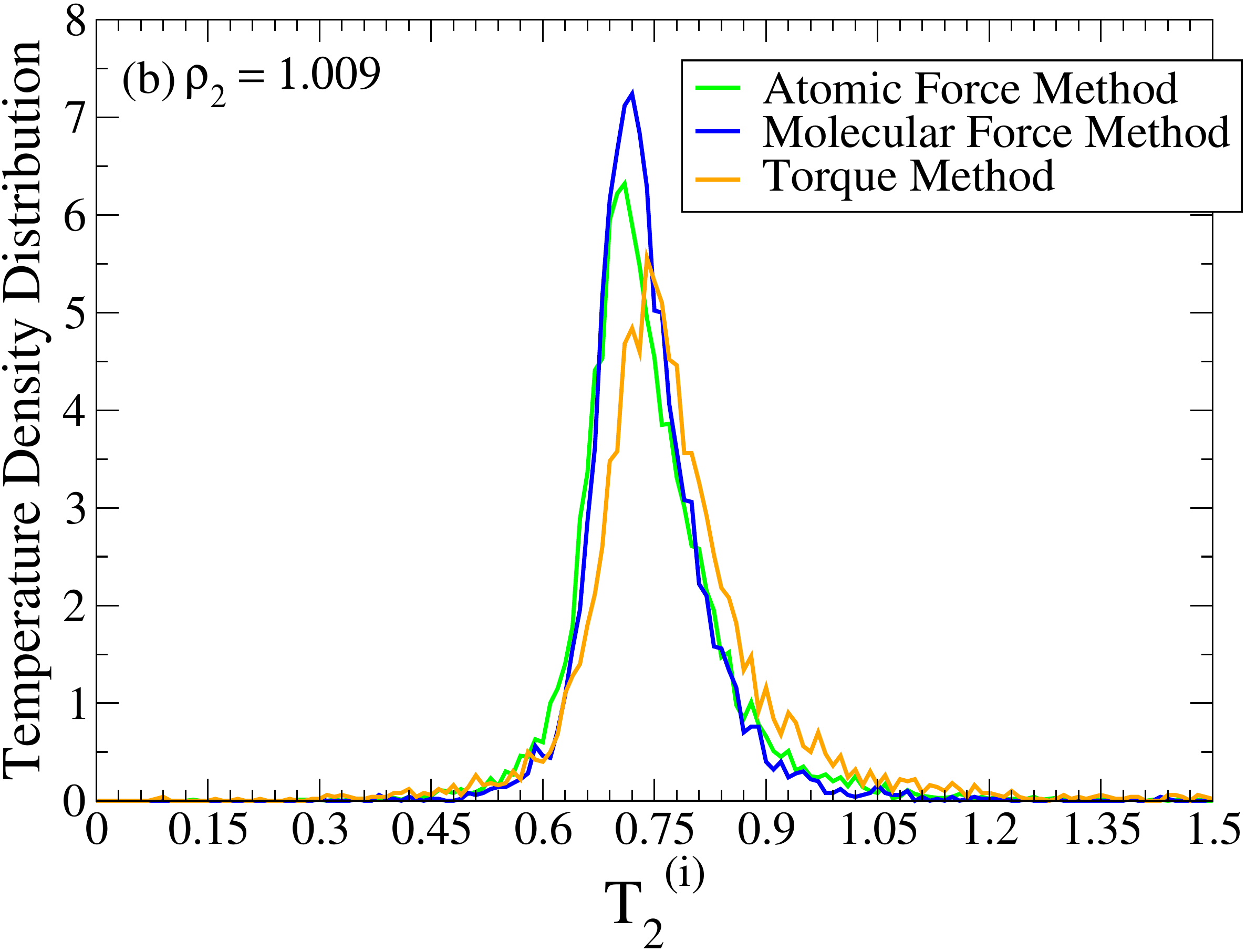}		
	\includegraphics[width=5.4cm]{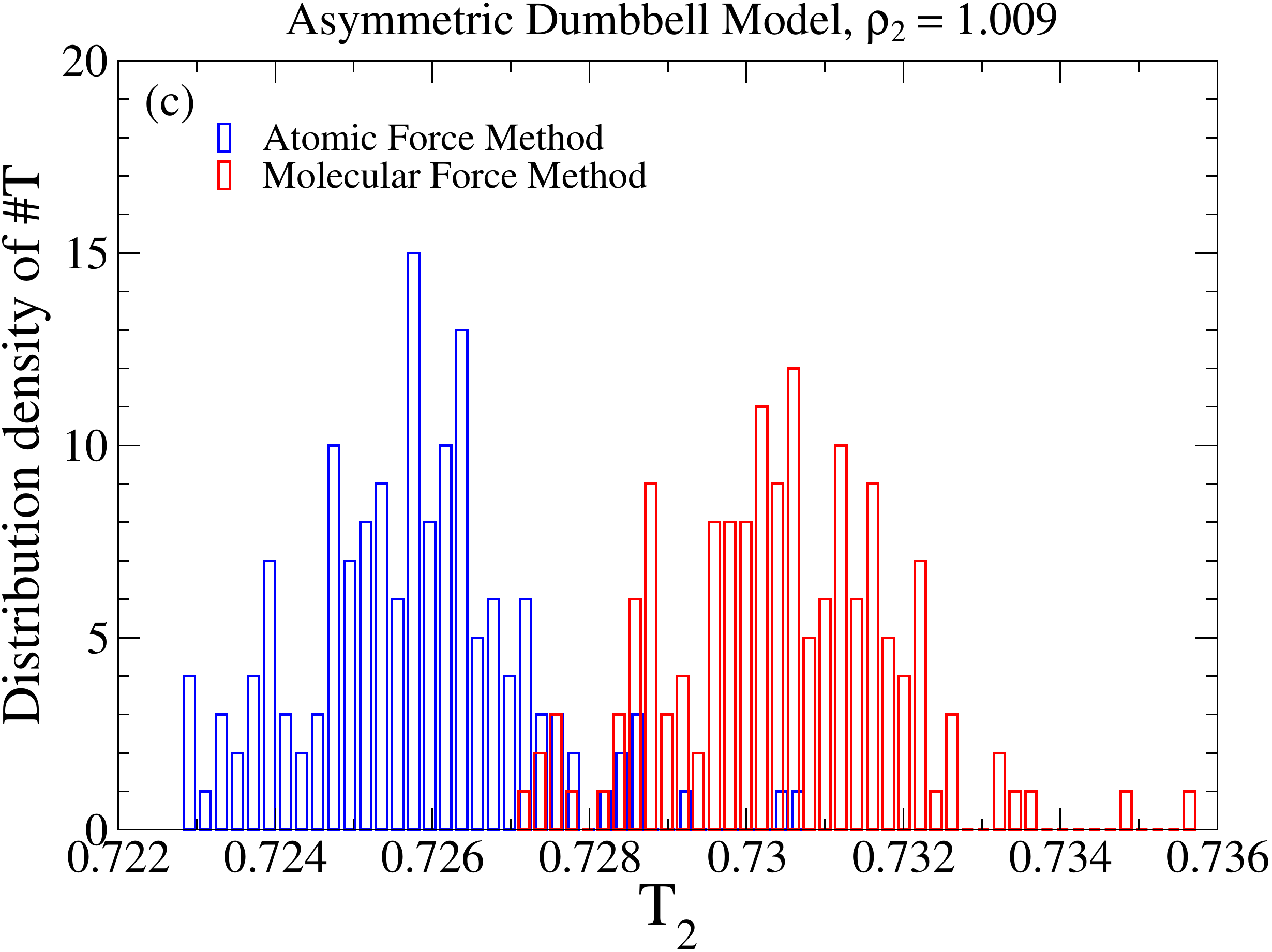}
	\caption{\label{FIG5}
		(a) Distribution of temperatures predicted by applying \eq{eq:pre_temp} and \eq{eq:pre_tor} to the individual molecules of a single configuration. For perfect scaling, all $T_2$ values should agree, i.e., the distributions should be identical delta functions.
		(b) The distribution of single-configuration $T_2$ values predicted from 152 independent configurations using the atomic-force (blue) and molecular-force (red) methods.
	}
\end{figure}

\Fig{FIG5}(a) shows the distribution of $T_2$ predictions when \eq{eq:pre_temp} and \eq{eq:pre_tor} are applied to the forces and torques of the individual molecules instead of to the high-dimensional configuration vector. The widths of the three distributions are similar.  \Fig{FIG5}(b) shows the distribution of $T_2$ values predicted by applying \eq{eq:pre_temp} to 152 independent configurations. Although scaling only a single configuration is a major advantage of the force- and torque-based methods, for comparing the methods reliably we report below average $T_2$ values predicted from 152 independent configurations.

\section{Comparing the three single-configuration methods}

\begin{figure}[htbp!]
	\centering
	\includegraphics[width=14cm]{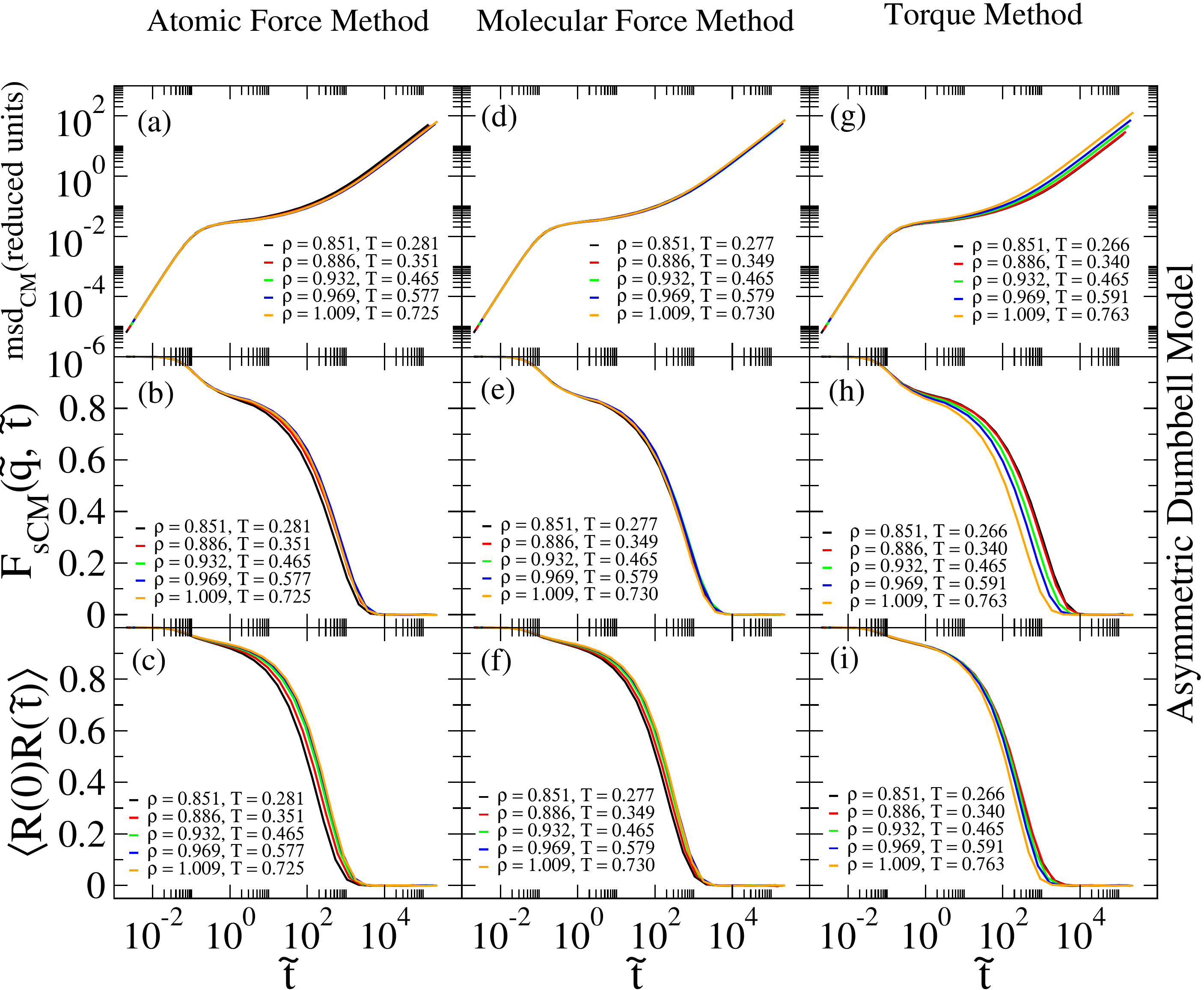}  
	\caption{\label{FIG6}Using the ASD model to test for invariance of the reduced translational and rotational dynamics by the three single-configuration methods. Each method investigates the reduced center-of-mass mean-square displacement (upper panels), center-of-mass incoherent intermediate scattering function ($\tilde q = 7.37$, middle panels), and orientational time-autocorrelation function of the normalized bond vector (lower panels). 
		(a), (b), (c) show results for state points generated by the atomic-force method requiring invariant reduced forces between all particles, thus including the intramolecular forces (\eq{eq:pre_temp}).
		(d), (e), (f) show results for state points generated by the molecular-force method requiring invariance of the reduced center-of-mass forces between the molecules (\eq{eq:pre_temp}).
		(g), (h), (i) show results for state points generated by the torque method requiring invariant reduced torques (\eq{eq:pre_tor}).
    Overall, the best invariance is obtained by the molecular-force method although the torque method works best for the short-time rotational dynamics.
	}
\end{figure}

We proceed to test the above methods on the three molecular models for invariance of the reduced-unit center-of-mass mean-square displacement (MSD), intermediate incoherent scattering function $F_s(q,t)$, and orientational time-autocorrelation function $\langle R(0)R(t)\rangle$.

\begin{figure}[htbp!]
	\centering
	\includegraphics[width=8cm]{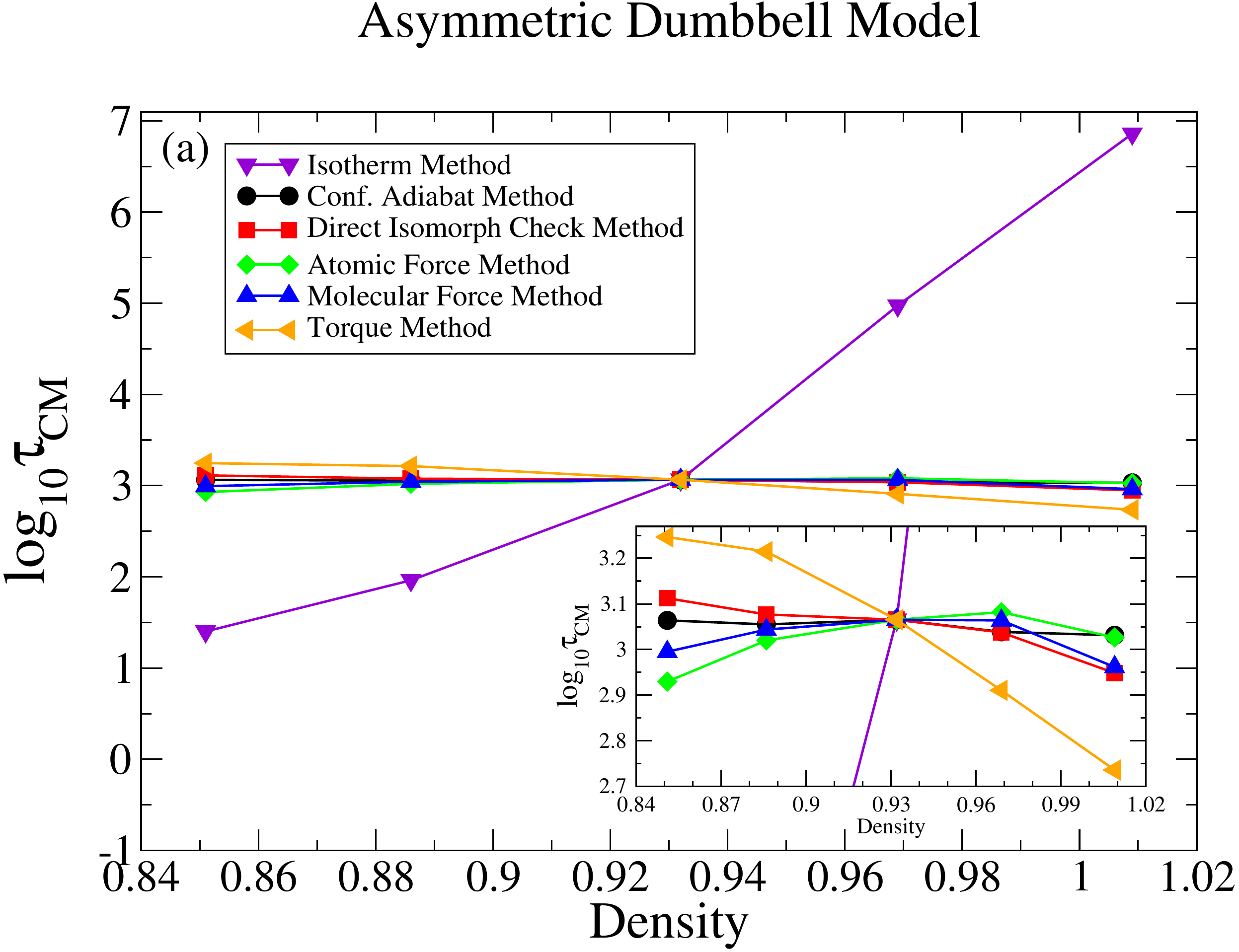}
	\includegraphics[width=8cm]{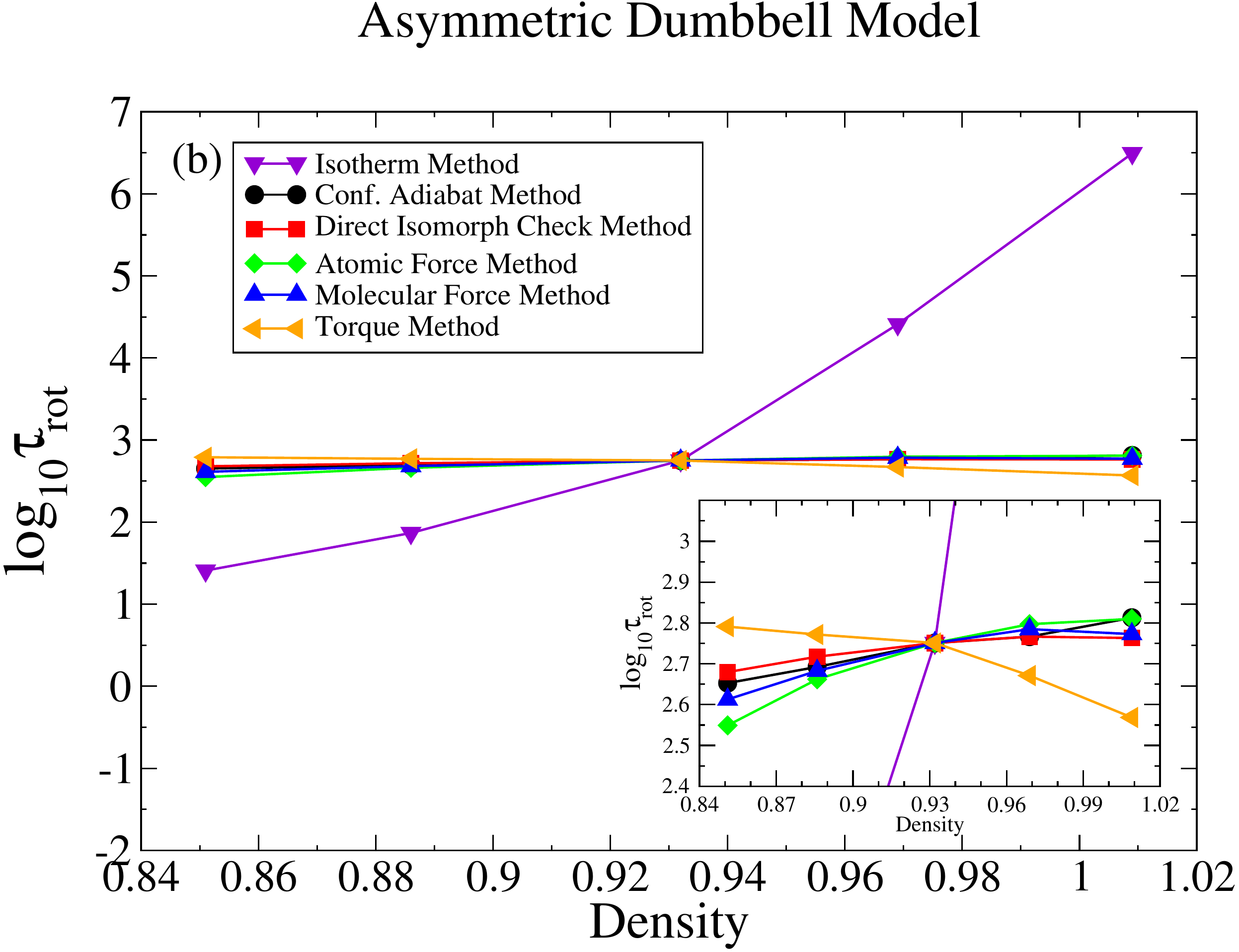}
	\caption{\label{FIG7} Comparing the relaxation time as a function of density for the ASD model along an isotherm (purple) and approximate isomorphs generated by the following five methods: configurational adiabat (black), direct isomorph check (red), atomic (green) and molecular (blue) force methods, and torque method (orange).
    (a) shows the reduced translational relaxation time calculated from the intermediate scattering function. 
    (b) shows a similar plot for the reduced rotational relaxation time calculated from the orientational time-autocorrelation function.}  
\end{figure}

\Fig{FIG6} shows results from the application of the three methods to the ASD model using $(\rho_1, T_1) = (0.932, 0.465)$ as reference state point. From this we determined two state points with lower density and two with higher density, spanning in all a density variation of 19\%. For the translational dynamics (MSD; upper row) the best invariance is obtained by the molecular-force method. The same applies for the incoherent intermediate scattering function, while the torque method is marginally better for the orientational time-autocorrelation function.

A more detailed analysis of the dynamics is given in \fig{FIG7} and Table \ref{tab_asy}. \Fig{FIG7} shows the  variation of the relaxation times of (a) the translational and (b) the rotational motion along an isotherm (purple), a configurational adiabat ($\Sex=$ Const., black), and curves generated by the direct-isomorph-check (red), atomic-force (green), molecular-force (blue), and torque methods (orange). The relaxation times reported here are defined from when the associated relaxation function has decayed to 0.2. All methods for tracing out isomorphs result in reduced relaxation times that, not surprisingly, are much more invariant than along the isotherm. For a quantitative comparison, Table I shows the density variation of the reduced diffusion coefficient and  the translational and rotational reduced relaxation times along the isotherm and the five approximate isomorphs. The diffusion coefficient is calculated from the diffusive part of the MSD (compare \fig{FIG6}). While the best invariance is obtained by the computationally demanding configurational-adiabat method, among the three single-configuration methods the best overall results are obtained by the molecular-force method.

\begin{table}[htbp!]
	\caption{\label{tab_asy}ASD-model density variation of the reduced diffusion coefficient (first row), relaxation time of the center-of-mass dynamics (second row), and relaxation time of the rotational dynamics (third row). Logarithmic derivatives are estimated from fitting the density range from 0.886 to 0.969. The associated uncertainty on the last digit is given in parenthesis. Not surprisingly, the logarithmic derivatives on the isotherm are large. The remaining columns represent, respectively, the configurational-adiabat ($\Sex=$ Const. traced out by \eq{gamma}), direct-isomorph-check (DIC), atomic- and molecular-force, and torque methods. The best invariance is obtained by the configurational-adiabat method, while the best of the single-configuration methods is the molecular force method. 
 }
	\begin{tabular}{ccccccc}
		\hline
		&Isotherm  & \eq{gamma} & DIC & $F_{Atom}$ & $F_{Mol}$ & Torque\\
		\hline
		$\dfrac{\partial log \tilde{D}}{\partial log \rho}$ & -70(2) & -0.5(4) & 1.1(4) & -1.4(2) & -0.9(4) & 7.47(6) 
		\\ \hline
		$\dfrac{\partial \log \tilde{\tau}_{cm}}{\partial \log \rho}$ & 77(3) & -0.4(1)  & -1.0(1) & 1.60(7) & 0.5(1) & -7.8(1) 
		\\ \hline 
		$\dfrac{\partial \log \tilde{\tau}_{rot}}{\partial \log \rho}$ & 65(3) & 1.9(1) & 1.26(2) & 3.47(3) & 2.62(7) & -2.6(2)
	\end{tabular}
\end{table}

\begin{figure}[htbp!]
	\centering
	\includegraphics[width=14cm]{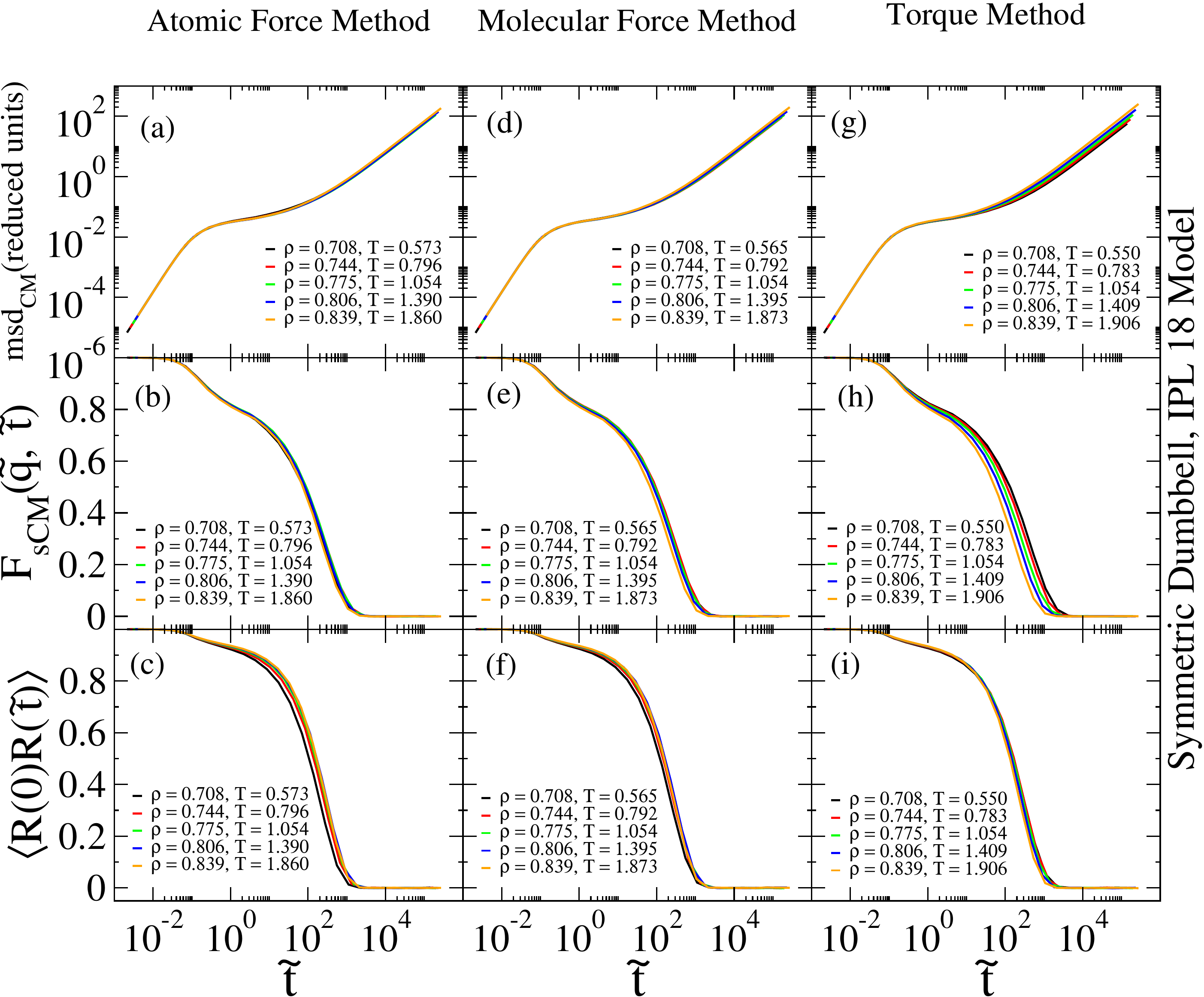}  
	\caption{\label{FIG8}Testing the atomic-force, molecular-force, and torque methods for the IPL model. The same dynamic quantities are investigated as in \fig{FIG6}. The reference point is $(\rho_1, T_1) = (0.775, 1.054)$ and $\tilde{q} = 7.97$. 
	(a), (b), (c) show results for state points generated by the atomic-force method.
	(d), (e), (f) show results for state points generated by the molecular-force method.
	(g), (h), (i) show results for state points generated by the torque method.
    Overall, the best results are obtained by the molecular-force method.	}
\end{figure}

\begin{figure}[htbp!]
	\centering
	\includegraphics[width=8cm]{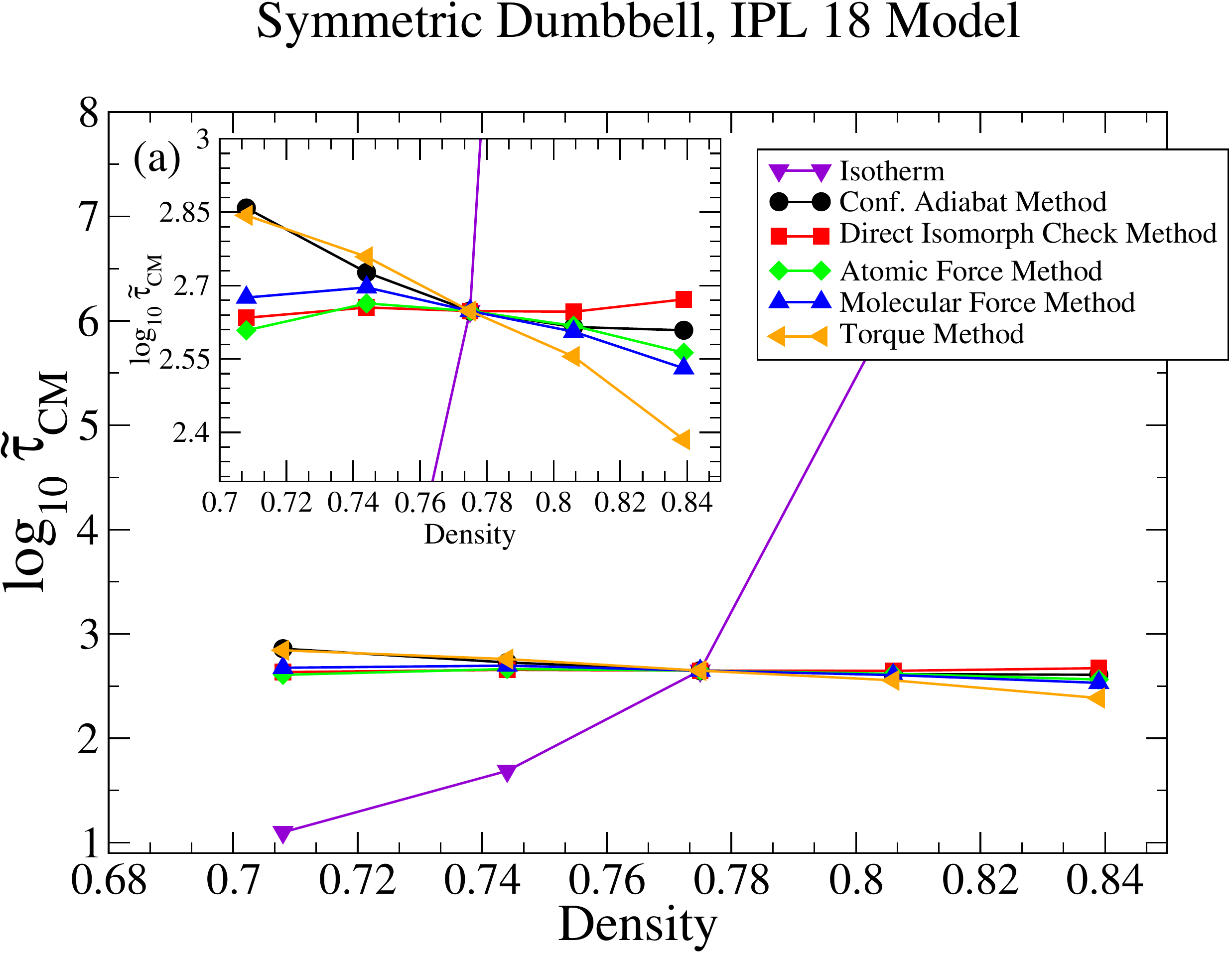}
	\includegraphics[width=8cm]{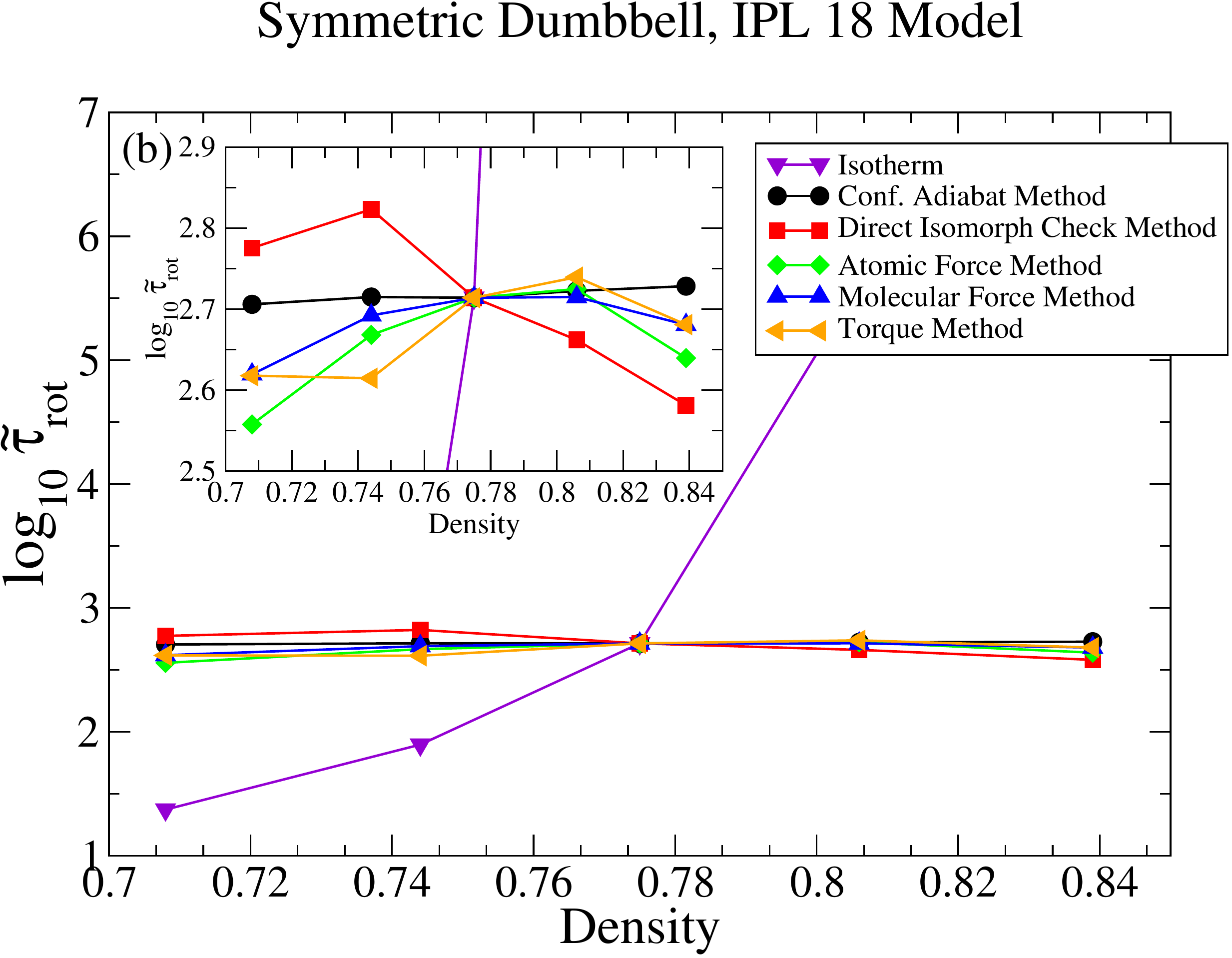}
	\caption{\footnotesize Comparing the reduced relaxation time as a function of density for the IPL model along an isortherm and the curves of \fig{FIG8}. As demonstrated previously, the approximate isomorph methods are all much better at obtaining invariant relaxation times than considering the same density variation along the isotherm. 
	(a) shows the reduced translational relaxation time calculated from the intermediate scattering function. 
	(b) shows a similar plot for the reduced rotational relaxation time.}  
	\label{FIG9}
\end{figure}

An important question is whether the  molecular geometry determines which method works best for which model. To investigate this we simulated the IPL symmetric dumbbell model (\fig{FIG8}, \fig{FIG9}, and Table \ref{tab_sym}). As for the ASD model, starting from the reference state point we applied each of the three single-configuration methods to generate two state points with lower density and two with higher density, spanning in all a density variation of 19\%. We find that the molecular-force method gives the best invariance curves for the rotational dynamics while the atomic-force method is best for the translational dynamics. \Fig{FIG9}(a) and (b) show the density variation of the translational and rotational dynamics by plotting their reduced relaxation times as in \fig{FIG7}.

\begin{table}[htbp!]
	\centering
	\caption{\label{tab_sym}Checking the reduced-units variation of the same dynamic quantities as in Table \ref{tab_asy} for the symmetric dumbbell IPL model. 
    Logarithmic derivatives are estimated from fitting the density range from 0.744 to 0.806. The associated uncertainty on the last digit is given in parenthesis. }
	\begin{tabular}{ccccccc}
		\hline \hline 
		&Isotherm  &  \eq{gamma} & DIC & \(F_{Atom}\) & \(F_{Mol}\) & Torque\\
		\hline
		\(\dfrac{\partial log \tilde{D}}{\partial log \rho}\) & -113.4(6) & 1.9(1) & -0.78(7)  & -0.2(2) & -0.462(5) & 3.9(3)\\
		\hline
		\(\dfrac{\partial \log \tilde{\tau}_{cm}}{\partial \log \rho}\)  & 126.7(7) & -0.4(3)  & -0.04(2) & -0.21(7) & -0.42(1) & -0.95(4) \\ 
		\hline 
		\(\dfrac{\partial \log \tilde{\tau}_{rot}}{\partial \log \rho}\) & 107.9(6) & 0.16(1) & -0.7(2) & 0.2(2) & 0.10(9) & 0.5(3)\\
		\hline
	\end{tabular}
\end{table}

\begin{figure}[htbp!]
	\centering
	\includegraphics[width=14cm]{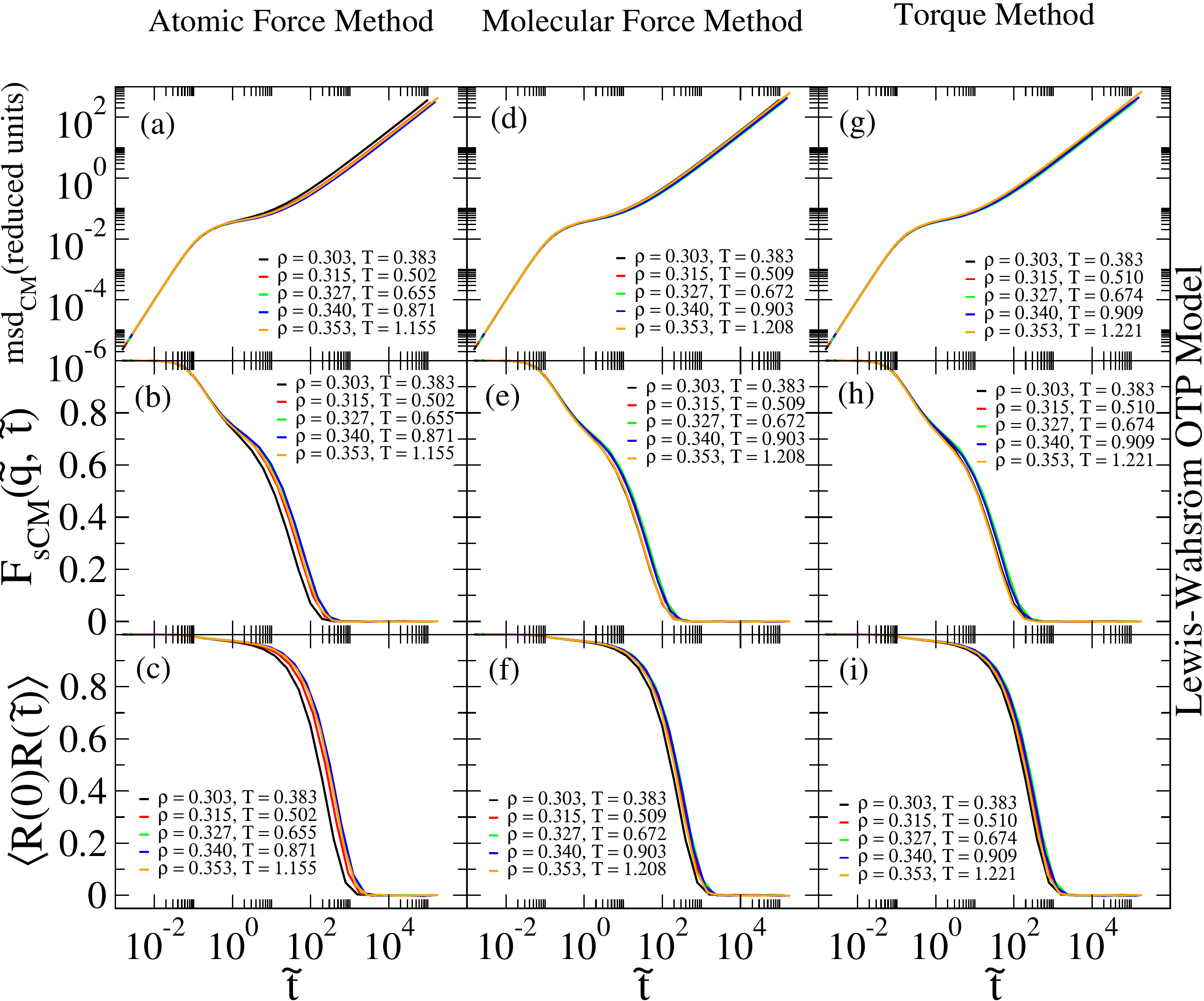}  
	\caption{\label{FIG10}Testing for invariance of the same reduced dynamic quantities as in \fig{FIG6} and \fig{FIG8} for the OTP model. The reference state point is $(\rho_1, T_1) = (0.303, 0.383)$ and $\tilde q = 10.2$. 
    (a), (b), (c) show results for state points generated by the atomic-force method.
    (d), (e), (f) show results for state points generated by the molecular-force method.
    (g), (h), (i) show results for state points generated by the torque method.
    The best results are obtained by the molecular-force and torque methods.}
\end{figure}

\begin{figure}[htbp!]
	\centering
	\includegraphics[width=8cm]{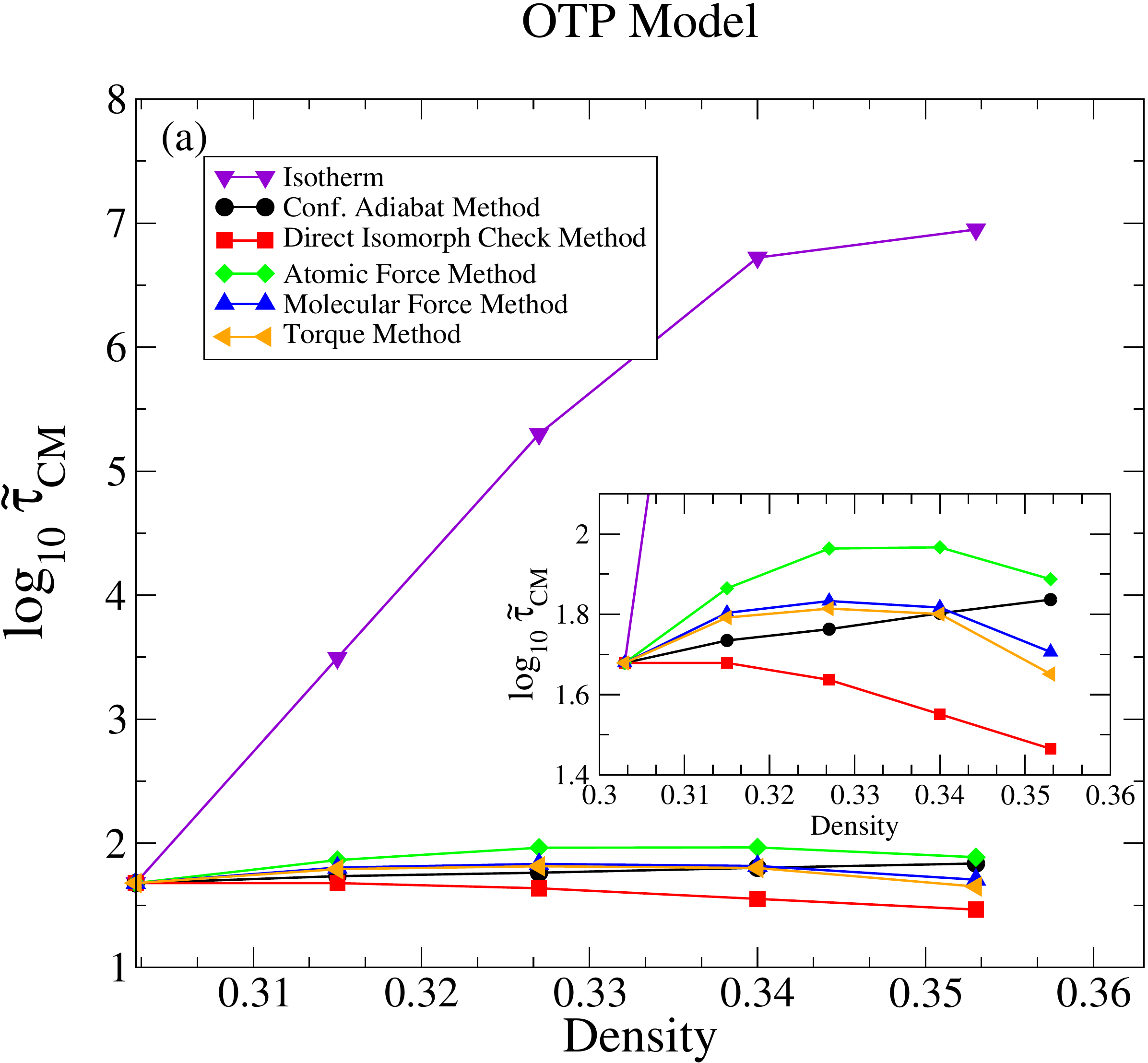}
	\includegraphics[width=8cm]{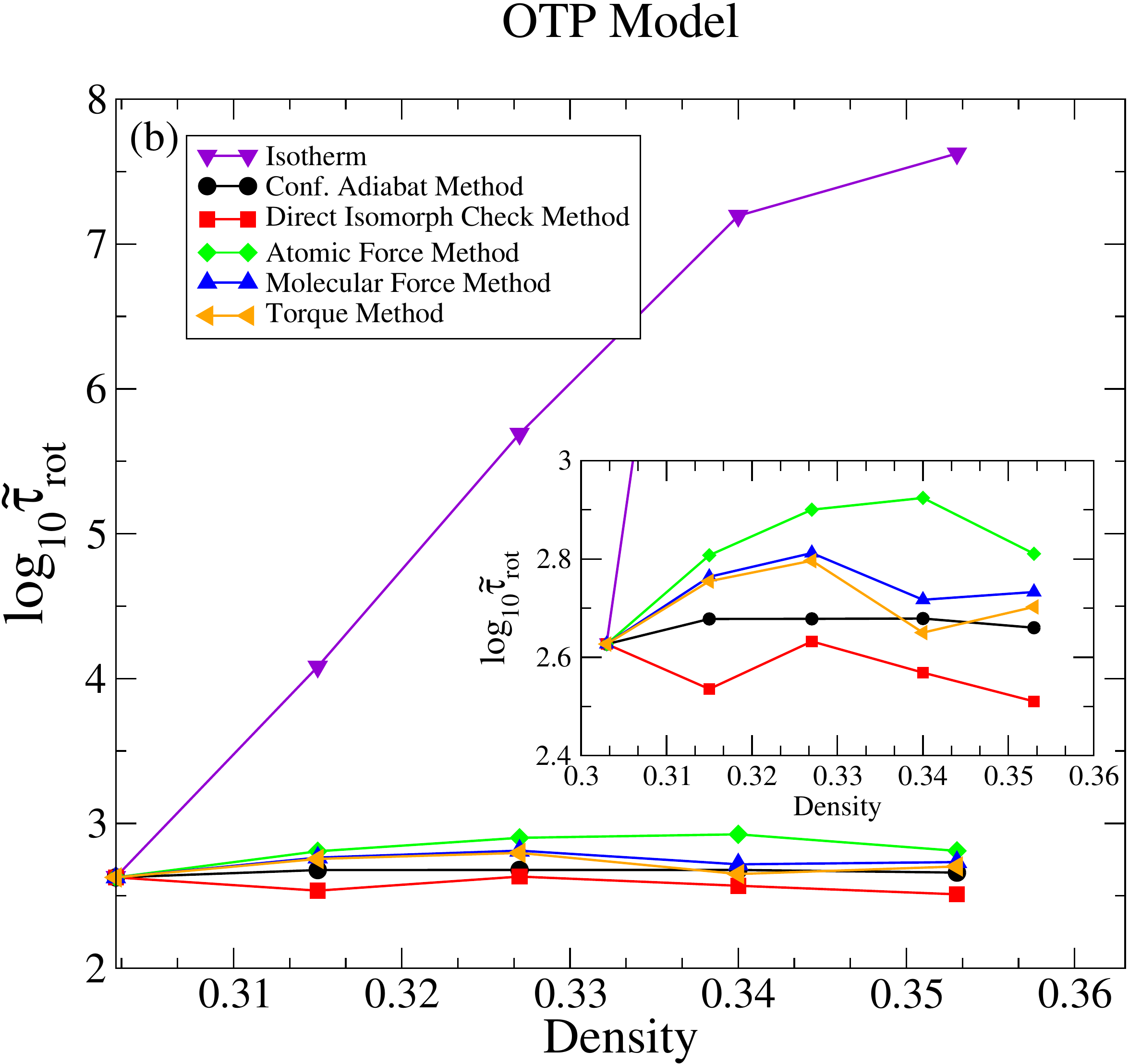}
	\caption{\footnotesize Comparing the reduced relaxation time as a function of density for the OTP model along an isotherm and the five methods for tracing out isomorphs. 
	(a) shows the translational relaxation time calculated from the intermediate scattering function.
        (b) shows a similar plot for the rotational relaxation time.}  
	\label{FIG11}
\end{figure}

We proceed to investigate the three single-configuration methods for the OTP model. \fig{FIG10} and \fig{FIG11} shows results using $(\rho_1, T_1) = (0.303, 0.383)$ as reference point, increasing from it the density by 16\%. In this case the molecular-force and torque methods give equally good invariance of the reduced-unit dynamics.

\begin{figure}[htbp!]
	\centering
 	\includegraphics[width=8cm]{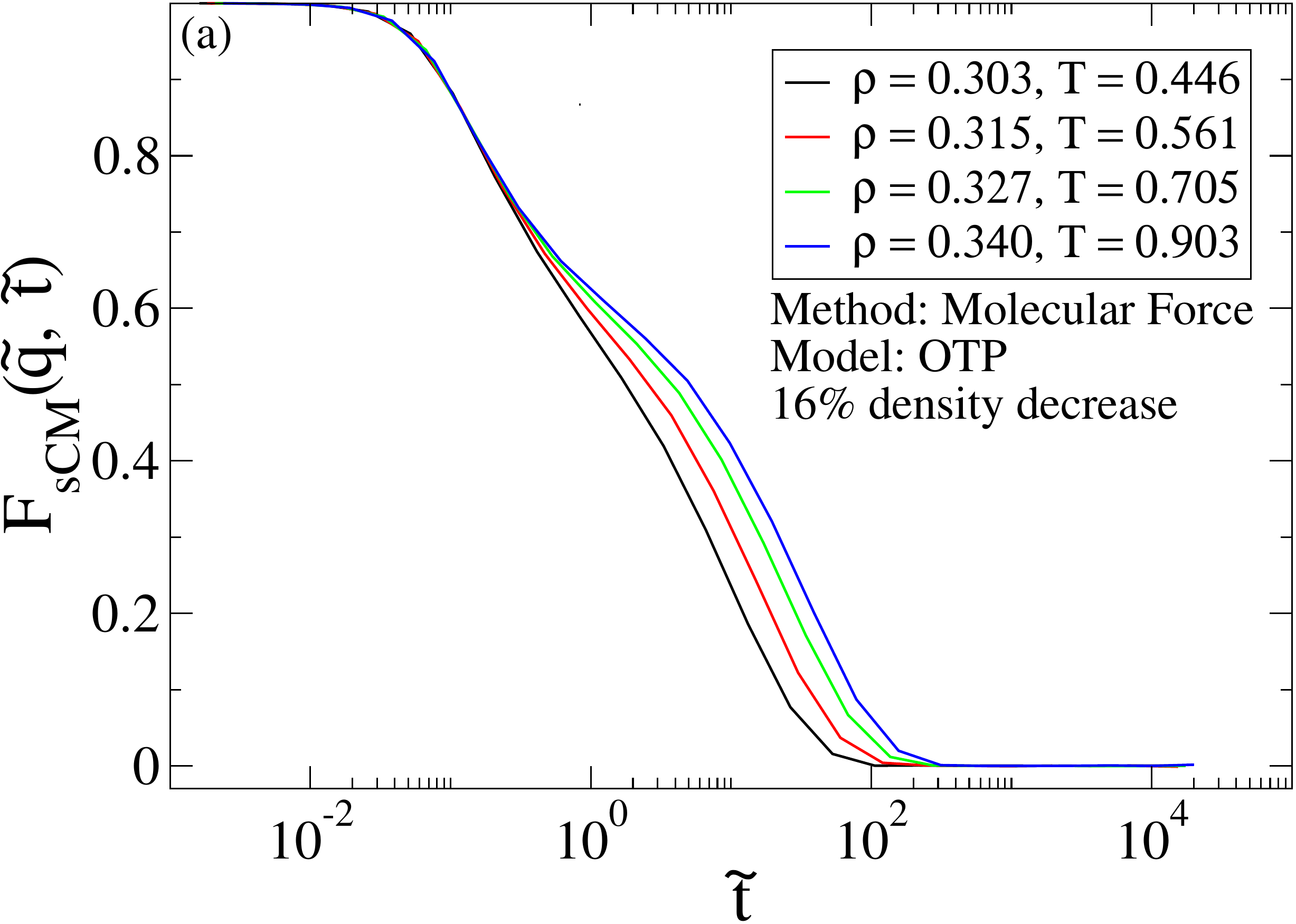}
        \includegraphics[width=8cm]{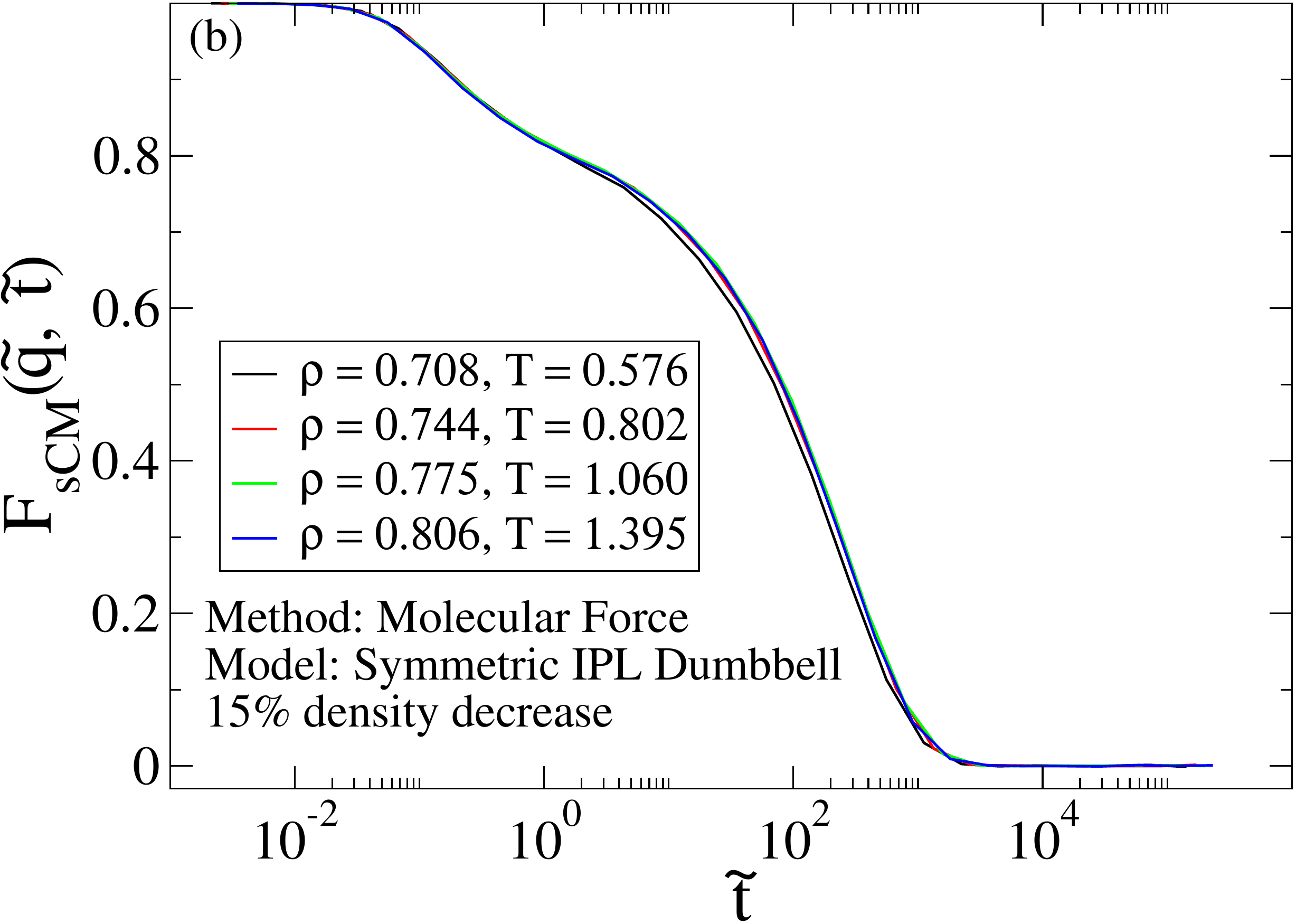}
	\caption{\footnotesize Comparing the dynamics of the OTP and IPL models when an isomorph is traced out by decreasing density from a reference state point using the molecular-force method.
    (a) gives results for the OTP model when the reference state point is $(\rho_1, T_1) = (0.340, 0.903)$, for the incoherent intermediate scattering function. In this case the dynamics is not invariant.   
    (b) tests the same for the IPL model using as reference state point $(\rho_1, T_1) = (0.806, 1.395)$ in which case there is still a good data collapse.}
	\label{FIG12}
\end{figure}

\begin{figure}[htbp!]
	\centering
	\includegraphics[width=8cm]{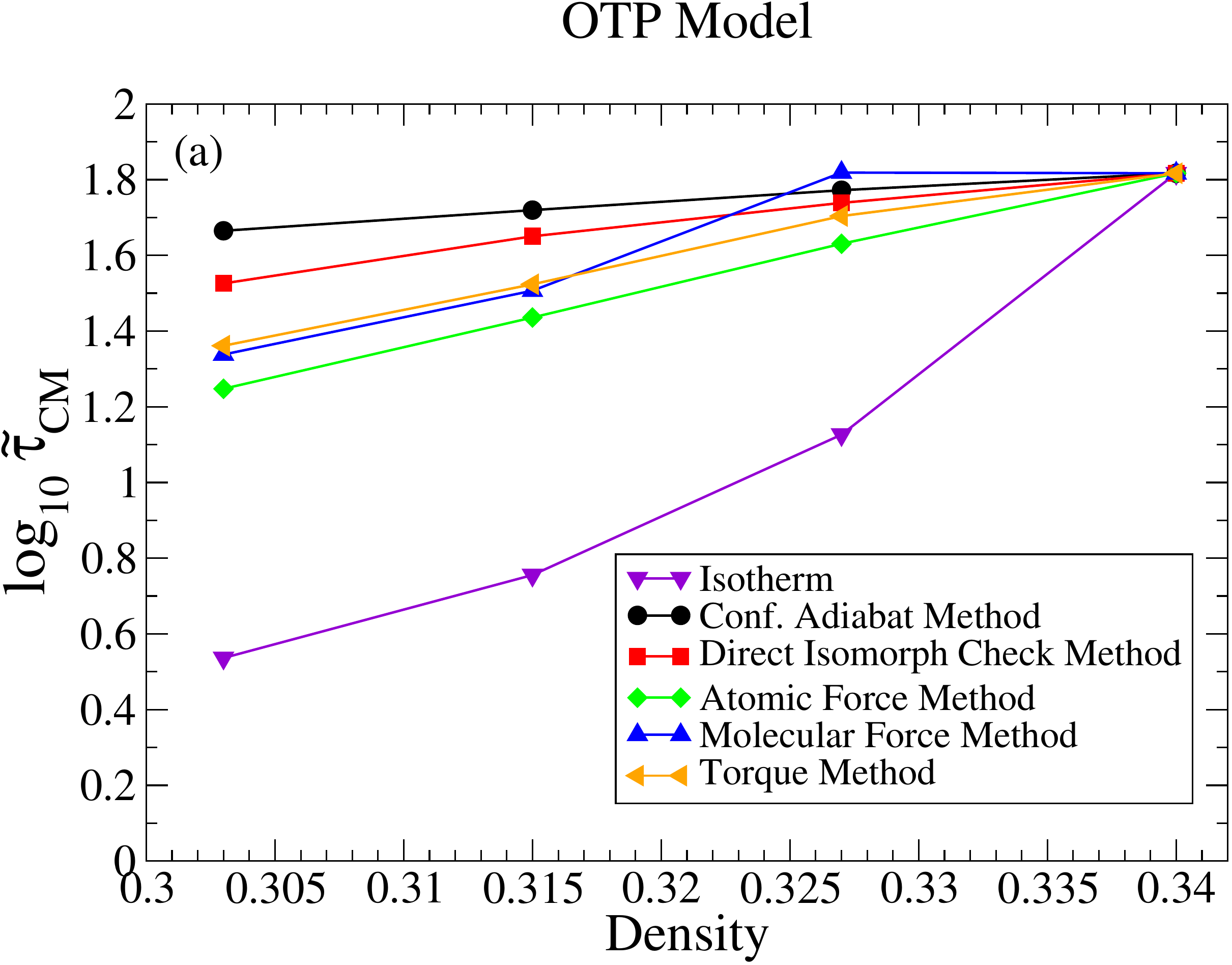}
	\includegraphics[width=8cm]{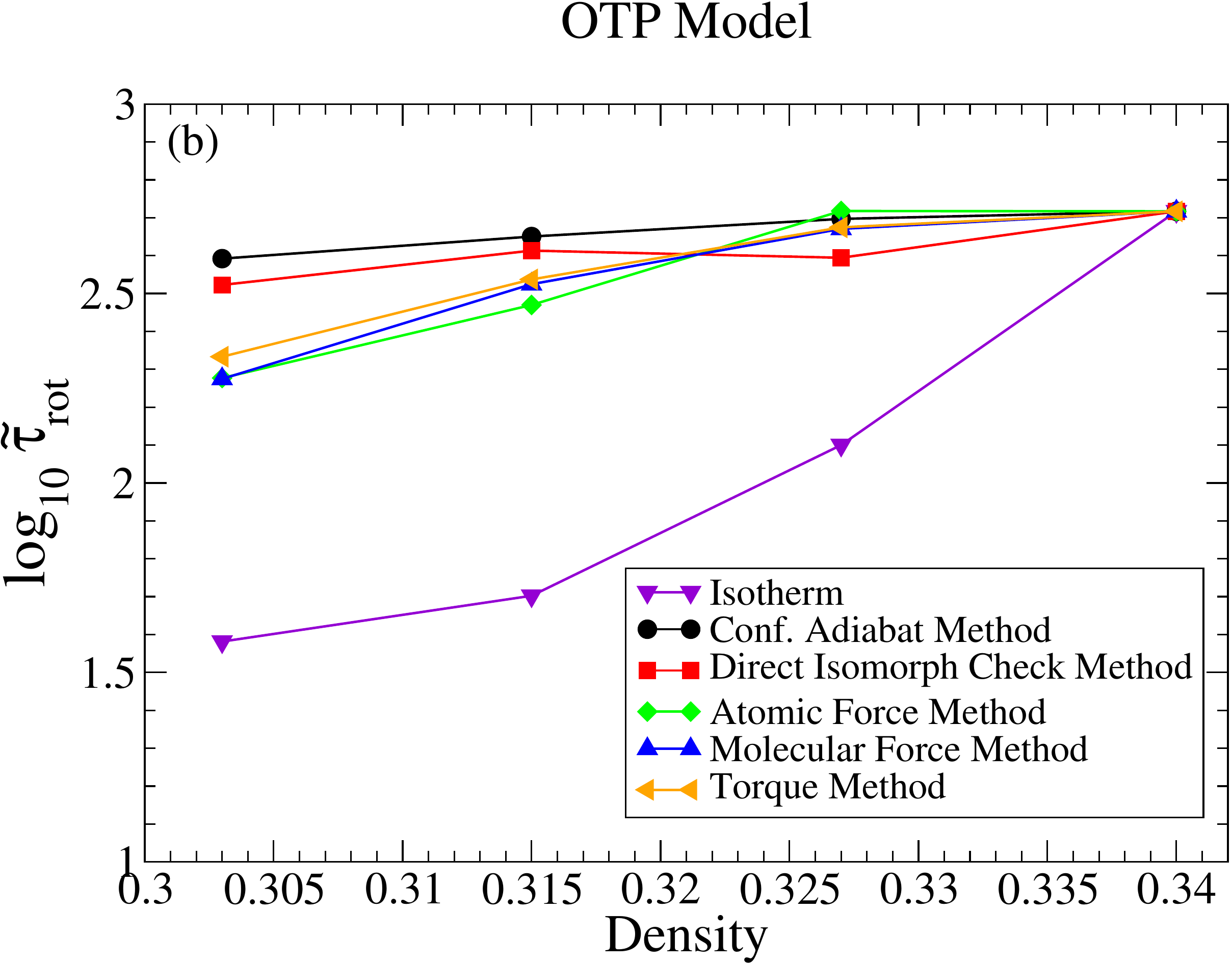}
	\caption{\footnotesize Variation of the reduced relaxation times of the OTP model when approximate isomorphs are generated using the molecular-force method by decreasing density from the reference state point. Again the reduced relaxation time is much more invariant along the different approximate isomorphs than along the isotherm. 
    (a) shows the translational relaxation time as a function of density. 
    (b) shows the same for the rotational dynamics. In both cases the dynamics is more invariant along the configurational-adiabat and direct-isomorphs-check generated isomorphs than along the isomorphs of the three single-configuration methods.}
	\label{FIG13}
\end{figure}

So far the results presented are quite similar for the three models tested; all three single-configuration methods work well, with the molecular force-method as the one that works best overall. However, our tests of the OTP model contains a surprise: how well the two force-based methods work depends on whether one increases or decreases the density. In the OTP tests reported above in \fig{FIG10} and \fig{FIG11}, density was \emph{increased} from the reference density, $\rho_1=0.303$, leading to invariant dynamics to a good approximation. In \fig{FIG12}(a) we take the fourth of the state points of \fig{FIG10} (e),  $(\rho_1, T_1) = (0.340, 0.903)$, as reference and move from there to lower densities. In this case, the intermediate scattering function is no longer invariant. Interestingly, this is only an issue with the OTP model. For the IPL model the dynamics is still invariant in reduced units when scaling to lower densities using the molecular-force methods: \Fig{FIG12}(b) shows the IPL model reduced incoherent intermediate scattering function of isomorphic points generated by starting from the reference state point $(\rho_1, T_1) = (0.806, 1.395)$ and from there decreasing the density.

\begin{figure}[htbp!]
	\centering
	\includegraphics[width=8cm]{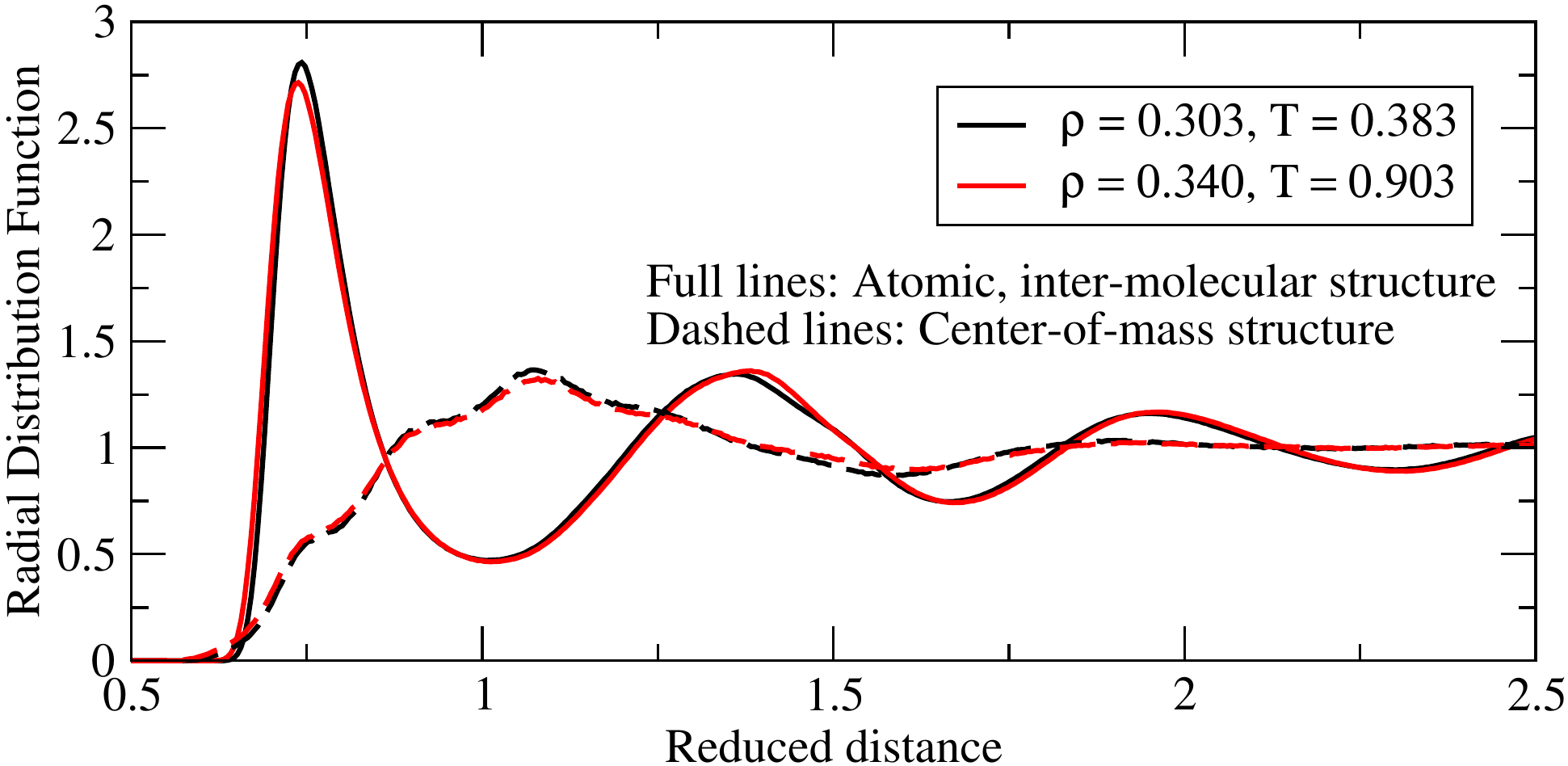}
	\caption{\footnotesize  Reduced unit structure for OTP at two state points discussed in the text. 
        Full lines: Reduced radial distribution function for the atoms, excluding intra-molecular contributions. 
        Dashed lines: Reduced radial distribution function of the center-of-masses of the molecules.
 }
	\label{rdfs}
\end{figure}

\fig{FIG13} shows the translational and rotational reduced relaxation times obtained by decreasing density for the OTP model from the reference state point $(\rho_1, T_1) = (0.340, 0.903)$. For the following discussion we focus on the two densities $0.303$ and $0.340$. Choosing $(\rho_1, T_1) = (0.303, 0.383)$ as reference state point, the molecular-force method gives $(\rho_2, T_2) = (0.340, 0.903)$ (\fig{FIG10} (e)). Choosing instead this as reference state point, $(\rho_1, T_1) = (0.340, 0.903)$ and decreasing density to the original value resulting in $(\rho_2, T_2) = (0.303, 0.446)$ (\fig{FIG13} (a)), i.e., a significantly different temperature than that of the original state point. How is this asymmetry possible? After all, \eq{eq:pre_temp} looks symmetric, so should one not expect to get back to the original reference state point? The important point here is that \eq{eq:pre_temp} is in fact \textit{not} symmetric: $\bRet$ is an equilibrium configuration at $(\rho_1, T_1)$, whereas $\bRto$ not necessarily is an equilibrium configuration at $(\rho_2, T_2)$. Instead $\bRto$ is an \emph{approximation} to an equilibrium configuration at the new state point, calculated by $\bRto = \left(\rho_1/\rho_2 \right)^{1/3}\bRet$. If the scaling were perfect, i.e., if \eq{Boltz} was exact, $\bRto$ would indeed be an equilibrium configuration. As discussed above, however, the scaling is seldom perfect, and it is not be perfect for the molecular systems studied here. This point is illustrated by \fig{rdfs} which shows the reduced-unit radial distribution functions at $(\rho, T) = (0.303, 0.383)$ and $(\rho, T) = (0.340, 0.903)$. While the reduced-unit structure is roughly invariant, this is clearly no exact. We believe this minor deviation explains the asymmetry between increasing and decreasing the density from the reference state point.

The arguments given above shows that it is possible for the force method to give different results depending on whether density is increased or decreased. At present we do not have a good explanation for why the molecular-force method works well when increasing the density and much worse when decreasing the density -- or why this effect is only seen for the OTP model. These issues deserve to be investigated in more detail in future works.

\section{Discussion and Conclusions}

Isomorphs exist in systems with strong virial potential-energy correlations, which include some systems of molecules with rigid bonds. For the ASD and IPL dumbbell and the Lewis-Wahnstr\"{o}m OTP models, we have seen that these curves to a good approximation can be traced out by methods based on a single configuration. Though not in focus here, the structure is also invariant to a good approximation along the approximate isomorphs identified by these methods.

We tested three methods to generate an approximate isomorph starting from a given reference state point. These methods all involve, in principle, just a single configuration and its uniformly scaled version, although we here averaged over 152 configuration pairs to compare the different methods. The atomic forces are affected by the intramolecular interactions, which explains why the atomic-force method is not well suited for identifying state points of approximately invariant reduced dynamics. The molecular-force method based on invariant reduced center-of-mass forces generally works best, although the torque method gives equally good results for the OTP model. Moreover, for the ASD and IPL models the torque method provides the best results for the short-time rotational dynamics. Overall, however, the best single-configuration method is the molecular-force method.

Identifying isomorphs via the three new methods is simpler and computationally much cheaper than using the configurational-adiabatic or direct-isomorph-check methods \cite{IV}. The force methods for generating isomorphs have here only been tested on molecular systems composed of molecules with rigid  bonds. The question whether the molecular-force method also works well for other molecular system, e.g., with harmonic bonds, is important to investigate in future works. Another point it would be interesting to investigate is whether the single-configuration methods discussed in this paper may be improved by allowing for molecular reorientations after the uniform scaling of the center-of-masses.

\acknowledgments{This work was supported by the VILLUM Foundation's \textit{Matter} grant (16515).}

\section*{References}

\end{document}